\documentstyle[multicol,aps,prb]{revtex}

\renewcommand{\narrowtext}{\begin{multicols}{2} \global\columnwidth20.5pc}
\renewcommand{\widetext}{\end{multicols} \global\columnwidth42.5pc}

\input{epsf.tex}

\def\top#1{\vskip #1\begin{picture}(290,80)(80,500)\thinlines \put(
65,500){\line( 1, 0){255}}\put(320,500){\line( 0, 1){
5}}\end{picture}}
\def\bottom#1{\vskip #1\begin{picture}(290,80)(80,500)\thinlines \put(
330,500){\line( 1, 0){255}}\put(330,500){\line( 0, -1){
5}}\end{picture}}

\begin{document}

\draft

\preprint{IUCM96-030, IUHET 350} 

\title{Edge-Magnetoplasmon Wave-Packet Revivals in the Quantum Hall
       Effect}
 
\author{Ulrich Z\"ulicke,$^1$ Robert Bluhm,$^2$
        V.\ Alan Kosteleck\'y,$^1$ and A. H.\ MacDonald$^1$}

\address{$^1$Department of Physics, Indiana University, Bloomington,
         IN 47405 \\ $^2$Department of Physics, Colby College,
         Waterville, ME 04901}

\date{\today}

\maketitle
 
{\tightenlines
\begin{abstract}
The quantum Hall effect is necessarily accompanied by low-energy
excitations localized at the edge of a two-dimensional electron
system. For the case of electrons interacting via the long-range
Coulomb interaction, these excitations are edge magnetoplasmons. We  
address the time evolution of localized edge-magnetoplasmon
wave packets. On short times the wave packets move along the edge
with classical $E$ cross $B$ drift. We show that on longer times the
wave packets can have properties similar to those of the Rydberg
wave packets that are produced in atoms using short-pulsed lasers.
In particular, we show that edge-magnetoplasmon wave packets can
exhibit periodic revivals in which a dispersed wave packet
reassembles into a localized one. We propose the study of
edge-magnetoplasmon wave packets as a tool to investigate dynamical
properties of integer and fractional quantum-Hall edges. Various
scenarios are discussed for preparing the initial wave packet and
for detecting it at a later time. We comment on the importance of 
magnetoplasmon-phonon coupling and on quantum and thermal
fluctuations.
\end{abstract}
}
 

\narrowtext

\section{Introduction}

The quantum Hall (QH) effect occurs in two-dimensional (2D) electron
systems (ES) whenever the chemical potential has a discontinuity
which occurs at a magnetic-field-dependent density.\cite{ahmintro}
For isotropic 2D ES, charge gaps occur at integer and certain
non-integer but rational values of the filling factor $\nu :=
\frac{n}{B}\,\Phi_0$, where $\Phi_0 = hc/e $ is the magnetic-flux
quantum. For the most part, we will assume in this paper that the 2D
ES has a filling factor that is the inverse of an odd integer: $\nu
= 1/m$, with $m$ odd. When the QH effect occurs, the bulk of the
system is incompressible in the absence of disorder, and the only
gapless excitations are localized near the boundary of the finite QH
sample.\cite{ahm:prl:90} For $\nu =1/m$ and the case of a sharp edge,
{\it i.e.}, a confining potential that varies rapidly on a length
scale of the order of the magnetic length
$\ell=\sqrt{\hbar c /|e B|}$, the edge excitations are
expected\cite{ahm:braz:96} to be well-described by a {\em chiral
Luttinger liquid} (CLL) theory\cite{wen:prb:90,wen:prb:91a,wen:int:92}
with a single branch of chiral bosons. In this theory, the edge of a
two-dimensional electron system is thought of as a one-dimensional
electron gas,\cite{emery,sol:adv:79,fdmh:jpc:81} which can be studied
using bosonization techniques. The only low-lying excitations are
collective bosonic density waves. For sharp $\nu =1/m$ edges in the
quantum-Hall regime, it turns out\cite{ahm:prl:90,wen:prb:90} that
there is a single branch of bosons and that these are chiral,
{\it i.e.}, they occur with only one sign of wave vector. If a
long-range Coulomb interaction is present, the bosons are
called edge magnetoplasmons\cite{hel1:85,emplcaveat,vav-sam:jetp:88}
(EMP). It is possible to derive an expression for the energy
dispersion relation of edge magnetoplasmons starting from the
microscopic Hamiltonian for electrons moving in a strong magnetic
field and interacting via a 3D Coulomb interaction.\cite{uz-ahm} The
result, which is valid for both the disk and strip geometries, is
\begin{equation}\label{disperse}
\varepsilon^{\text{C}}_{M} = - \frac{M}{R} \ln{\left[\alpha
\frac{M}{R} \right]} \quad .
\end{equation}
Expressions differing from Eq.~(\ref{disperse}) only in the constant 
$\alpha$ had been obtained earlier using a semiclassical
approach.\cite{vav-sam:jetp:88,wass:prb:90,bla:phyb:92} Here, the
constant $\alpha$ is of order unity and depends weakly on the
geometry of the QH sample, $M$ is a positive integer, and $R$ is
related to the perimeter $L$ of the QH sample: $R={L}/{2\pi}$. For
the disk geometry, $R$ is the radius of the disk. In
Eq.~(\ref{disperse}), as well as in all expressions to follow in this
paper, we measure physical quantities in {\em quantum-Hall units} to
simplify expressions. These units are defined in
Table~\ref{table_nsu}.

We can think of the edge of a QH sample as an excitable
one-dimensional medium, much like a string. The edge magnetoplasmons
are the eigenmodes of this medium. For the case of a short-range
interaction between the electrons, their dispersion is linear in the
wave number: $\varepsilon^{\text{sr}}_{M} = v_{\text{F}} \, M / R$.
If a long-range Coulomb interaction is present, the dispersion
relation is nonlinear and is given by Eq.~(\ref{disperse}).
\begin{table}[hbt]
\caption{Quantum-Hall units. Throughout this paper, physical
quantities are measured in these units. Besides the formal expression
for each quantum-Hall unit, explicit values are given (as a function
of magnetic field $B$ and filling factor $\nu$) which apply to 2D
ES in GaAlAs/GaAs-heterostructures. The symbols $e$, $\hbar$, and
$k_B$ denote the electron charge, Planck constant and Boltzmann
constant, respectively.}
\begin{tabular}{rcc}
Quantity & QH unit (in cgs) & Values (for GaAs) \\ \hline
length & $\ell = \sqrt{\hbar c / |e B|}$ & $25.7 \times
B[\rm T]^{- 1/2} \,$~nm \\
energy & $(\nu e^2) / (\pi \epsilon \ell)$ & $1.39 \times \nu \,
B[\rm T]^{1/2}$~meV \\
time   & $(\hbar \pi \epsilon \ell) / (\nu e^2)$ & $4.74 \times
10^{-13} / \nu B[\rm T]^{1/2}$~s \\
temperature &$(\nu e^2) / (\pi \epsilon \ell k_B)$ & $16 \times
\nu B[\rm T]^{1/2}$~K \\
\end{tabular}
\label{table_nsu}
\end{table}
\noindent
Using a time-dependent external potential, it is possible to excite a
superposition of these eigenmodes. In this way, a wave packet can be
created for the electron number density along the edge. We refer to
these as edge-magnetoplasmon wave packets (EMP WP).

\begin{figure}[hbt]
\epsfxsize=8.5cm
\centerline{\epsffile{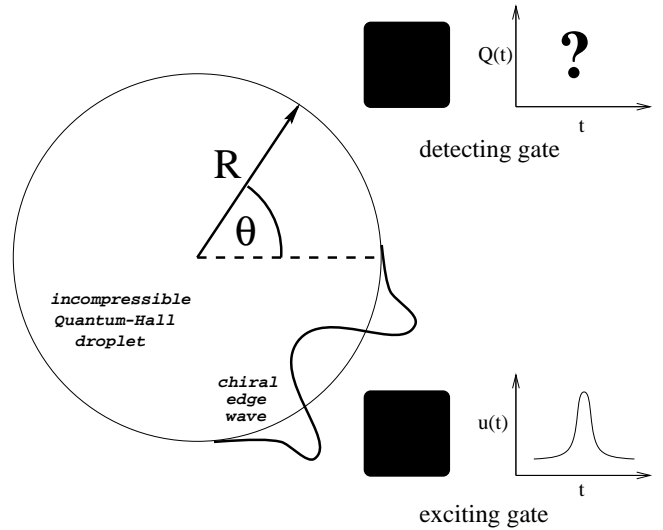}}
\caption{Schematic picture of a chiral-edge-wave excitation and
detection. A voltage pulse is applied to the exciting gate, leading
to the formation of an edge-magnetoplasmon wave packet, {\it i.e.}, a
superposition of edge-magnetoplasmon (EMP) modes. This wave packet
propagates according to the dynamics of EMP and can be detected at
a second gate. The study of the detected pulse provides the
opportunity to investigate the dynamical properties of EMP modes as
well as the revival structure of edge-magnetoplasmon wave packets.}
\label{geometry}
\end{figure}

In this paper, we examine the formation and evolution of EMP WP in
two-dimensional electron systems. Experimental
studies\cite{hls:prb:83,tal:jetp:89,heit:prl:90,
heit:prl:91,ray:prb:92,zhi:prl:93,hel2:85,hel:91,hel1:85,wass:prb:90}
of EMP have employed different excitation processes, including
capacitive coupling\cite{ray:prb:92} or ohmic
contacts\cite{zhi:prl:93} between the QH edge and an exciting voltage
pulse. This paper is motivated most strongly by the capacitive
coupling\cite{ray:prb:92} approach and we comment later on the
interpretation of these experiments. The physical observables
involved in these probes can all be expressed in terms of the
operators involving the creation and annihilation operators for
bosonic density fluctuations (EMP): $b_{M}^{\dagger}$ and $b_{M}$.
In particular, the one-dimensional number density of electrons at the
edge of a sample with filling factor $\nu$, defined by integrating
perpendicular to the edge and comparing with the ground-state density,
is related to boson creation and annihilation operators
by\cite{wen:int:92}
\begin{equation}
\label{density}
\varrho(\theta) = \sum_{M>0} \frac{(\nu M)^{1/2}}{2\pi R} \, \left[
b_{M} \, e^{i M \theta} + b^{\dagger}_{M} \, e^{- i M \theta} \right]
\quad .
\end{equation}
In this expression, we have fixed the direction of the magnetic field.
Reversing the field direction corresponds to interchanging
$b_{M}^{\dagger}$ and $b_{M}$. In our convention, the EMP travel
counterclockwise. The experimental configuration is shown in
Fig.~\ref{geometry}.

Localized electron wave packets have been produced and studied in
atoms using short-pulsed lasers.\cite{ps,az,tenWolde,yeazell,meacher}
A superposition of highly excited or Rydberg states is formed when a
short laser pulse coherently excites a single electron far from the
ground state of an atom. The resulting Rydberg wave packet is
localized spatially, and its initial motion mimics the classical
periodic motion of a charged particle in a Coulomb potential. The
wave packet will eventually disperse and lose its classical character.
However, the wave packet reassembles at later times in a sequence of
fractional and full revivals and superrevivals.\cite{ps,az,ap,nau,sr}
The revivals result from quantum interference between the different
eigenstates in the superposition. The appearance of revivals is quite
general and can occur in quantum systems other than Rydberg
atoms,\cite{bkp} including systems with eigenenergies depending on
more than one quantum number.\cite{revs23}

In the present work, we perform an analysis of the revival structure
of EMP WP in two-dimensional electron systems that on short times
exhibit classical $E$ cross $B$ drift motion in the electric field
that confines the electronic system to a finite area. At longer
times these wave packets disperse but we show that they can also
have fractional and full revivals and superrevivals.

We begin in Section~\ref{sec2} with a brief review of Rydberg
wave-packet behavior, including a discussion of fractional and full
revival of dispersed wave packets. Section~\ref{sec3} describes the
dynamics of EMP in QH samples with sharp edges. The exact solution of
the time-evolved edge-magnetoplasmon wave packet is obtained. We
demonstrate that EMP WP can exhibit full and fractional revivals that
we illustrate for one particular set of experimental parameters. The
similarities between Rydberg wave packets and EMP WP are examined. In
Section~\ref{sec4}, we discuss experimental issues, taking into
account the decay of EMP WP due to scattering processes occurring in
the semiconductor host material. Our conclusions are presented in
Section~\ref{sec5}. Some details of the derivations are given
separately in appendices.

\section{Atomic Rydberg Wave Packets}\label{sec2}

In this section, we present some background on atomic Rydberg
wave packets. These wave packets form when a short-pulsed laser
excites an atomic electron into a coherent superposition of
large-quantum-number Rydberg states. Several theoretical approaches
have been used to investigate Rydberg wave packets. They have been
studied both numerically\cite{ps} and perturbatively.\cite{az} A
description in terms of squeezed states has also been
given.\cite{squeezed} Rydberg wave packets offer the possibility of
approaching the classical limit of motion for electrons in atoms.
Initially, the motion of a Rydberg wave packet is semiclassical,
exhibiting the classical periodic motion of a charged particle in a
Coulomb field. The period of the motion is the classical keplerian
period $T_{\text{cl}}$. This semiclassical motion typically lasts for
only a few cycles, after which the wave packet disperses and
collapses. However, quantum interference effects subsequently cause
the wave packet to undergo a sequence of fractional and full revivals.

The full revivals are characterized by the recombination of the
collapsed wave packet into a form that resembles the initial
wave packet. This first occurs at a time $t_{\text{rev}}$. At the full
revival, the wave packet again oscillates with the classical period
$T_{\text{cl}}$. The fractional revivals occur at earlier times equal
to irreducible rational fractions of $t_{\text{rev}}$. At the
fractional revivals, the wave packet separates into a set of equally
weighted subsidiary wave packets. The motion of the subsidiary
wave packets is periodic with period equal to a rational fraction of
the classical period $T_{\text{cl}}$.\cite{ap}  Eventually the
periodic revivals fail, but on a still longer time scale they can
reappear and a higher level of revival structure commences.\cite{sr}
This occurs on a longer time scale called the superrevival time
$t_{\text{sr}}$. At times equal to certain rational fractions of
$t_{\text{sr}}$, distinct subsidiary waves form again, but with a
period equal to a rational fraction of $t_{\text{rev}}$. These
long-term fractional revivals culminate with the formation of a
superrevival at the time $t_{\text{sr}}$. At the superrevival, the
wave packet can resemble the initial wave packet more closely than at
the full revival time $t_{\text{rev}}$. An analysis including the
superrevival time scale has been performed\cite{sr} both for
hydrogenic models and using supersymmetry-based quantum-defect
theory,\cite{sqdt} to model the Rydberg alkali-metal atoms typically
used in experiments. For times typically a few orders of magnitude
greater than $t_{\text{sr}}$, atomic Rydberg wave packets
spontaneously decay into lower-energy states by emitting photons.

Experiments\cite{tenWolde,yeazell,meacher,wals} have studied the 
revival structure of Rydberg wave packets through times
$\sim t_{\text{rev}}$. These experiments use a pump-probe method of
detection involving either photoionization\cite{az} or
phase-sensitive Ramsey interference and electric-field
ionization.\cite{noord,broers,christian} In both these procedures,
the wave packet is excited initially by a short laser pulse that
creates a superposition of energy eigenstates with principal quantum
number centered on a value $\bar n$. The wave packet initially forms
near the nuclear core of the atom. After the pump pulse has passed,
the wave packet evolves under the influence of the Coulomb potential,
oscillating between inner and outer apsidal points of the keplerian
ellipse corresponding to $\bar n$. 

In pump-probe experiments using photoionization, a second laser pulse
called the probe pulse ionizes the atom. The photoionization signal
is measured as a function of the delay time $t$ between the pump and
probe signals. The transition probability for absorbing the second
photon is greatest when the wave packet is near the core and falls to
zero as the wave packet moves away from the nucleus. As a result, the
periodicity in the photoionization signal corresponds to the
periodicity in the probability for the wave packet to return to its
initial configuration. Experiments using this method have detected
the initial periodic motion of the wave packet with period
$T_{\text{cl}}$, as well as fractional revivals at delay times equal
to fractions of $t_{\text{rev}}$, and a full revival at
$t_{\text{rev}}$. The fractional revivals are characterized by
periodicities in the photoionization signals equal to rational
fractions of $T_{\text{cl}}$, as expected.

A second type of pump-probe experiment is based on Ramsey's method of
separated oscillating fields.\cite{norm} In this method, an initial
laser pulse creates a wave packet, which is then excited by a second
identical laser pulse. Depending on the relative phase between the
two time-separated optical pulses, the upper-state population in the
wave packet can be either increased or reduced by the second pulse.
This population transfer between the excited levels and the ground
state falls to zero as the wave packet moves towards its outer
turning point. The population of the excited levels as a function of
the delay time thus appears as a rapidly oscillating function, due to
the Ramsey interference, that is modulated by an envelope dependent
on the wave-packet motion. By monitoring the population of the
excited levels as a function of time using electric-field ionization,
the motion of the wave packet can be detected via the periodicities
in the envelope function. Since electric-field ionization is more
efficient than photoionization, the Ramsey method is better able to
resolve fractional revivals. Indeed, using this method, fractional
revivals consisting of as many as seven subsidiary wave packets have
been detected.\cite{wals}

The wave function for a Rydberg wave packet can be written as a
superposition of energy eigenstates
\begin{equation}\label{psi}
\Psi(\vec r,t) = \sum_n c_n \psi_n (\vec r) e^{-i E_n t / \hbar}
\quad ,
\end{equation}
where $c_n$ are weighting coefficients and $\psi_n$ are energy
eigenstates. For excitations with a short laser pulse, the
coefficients $c_n$ are strongly peaked around a central value of $n$
equal to $\bar n$. This permits a Taylor expansion of the energy
around the value $\bar n$. The first three derivative terms in this
expansion define the time scales $T_{\text{cl}}$, $t_{\text{rev}}$,
and $t_{\text{sr}}$ as follows:
\begin{mathletters}
\label{gentimes}
\begin{eqnarray}
T_{\text{cl}} &=& \frac {2 \pi} {\vert (E_{\bar n})^\prime \vert}
\quad , \\
t_{\text{rev}} &=&  \frac {2 \pi} {\frac 1 2 \vert (E_{\bar n}
)^{\prime\prime} \vert} \quad , \\  
t_{\text{sr}} &=& \frac {2 \pi} {\frac 1 6 \vert (E_{\bar n}
)^{\prime\prime\prime} \vert} \quad .
\end{eqnarray}
\end{mathletters}
The primes denote derivatives with respect to $n$. The evolution of
the wave packet is controlled by the interplay between these three
time scales in the time-dependent phase.

For short times $t \ll t_{\text{rev}}$, only the first-derivative term
in the Taylor expansion of $E_n$ contributes significantly, and we
can write $\psi(\vec r,t) \approx \psi_{\text{cl}}(\vec r,t)$, where
\begin{equation}
\label{psicl}
\psi_{\text{cl}}(\vec r,t) = \sum_n c_n \psi_n (\vec r)
e^{-i 2 \pi (n - \bar n)t / T_{\text{cl}} } \quad ,
\end{equation}
and we have omitted an overall phase. The function $\psi_{\text{cl}}$
is periodic with period $T_{\text{cl}}$. This expression is valid only
for $ t \ll t_{\text{rev}}$, however. Eventually the second-derivative
term in the phase becomes appreciable, and the initial periodic motion
is destroyed. The fractional revivals occur at times $t = m \,
t_{\text{rev}}/n$, where $m$ and $n$ are relatively prime integers
($m\le n$). At these times, it has been shown\cite{ap} that, with
third and higher derivatives neglected, the wave function can be
written as a sum of subsidiary wave functions:
\begin{equation}
\label{fracrevs}
\Psi(\vec r,t) \approx \sum_{s=0}^{l-1} a_s \psi_{\text{cl}}(\vec r,t
+ \frac s l T_{\text{cl}}) \quad .
\end{equation}
The coefficients $a_s$ and the integers $l$ depend on $m$ and $n$ and
are defined in Ref.~\onlinecite{ap}. The moduli of the $a_s$ are
equal for all $l$, which means that the terms in the sum are equally
weighted. The form of this equation shows that at the fractional
revivals, the wave function equals a sum of subsidiary waves, each of
which has the form of the initial wave function but is shifted in its
orbit by a fraction of $T_{\text{cl}}$.

The periodicities in the motion of the wave packet are most easily
studied using the autocorrelation function\cite{nau}
$A(t) = \left< \Psi(\vec r,0) \vert \Psi(\vec r,t) \right>$. The
absolute square of the autocorrelation function as a function of time
gives the probability for the wave packet to return to its initial
configuration. In a pump-probe experiment using photoionization, the
periodicities in the photoionization signal should match those in
$\vert A(t) \vert^2$, since both measure the probability for the
wave packet to return to the core. Alternatively, in pump-probe
experiments using the Ramsey method, the ionization signal depends on
the real part of the autocorrelation function $\Re e \{ A(t) \} =
\Re e \left\{ \left< \Psi(\vec r,0) \vert \Psi(\vec r,t) \right>
\right \}$. This highly oscillatory function is modulated by an
envelope that depends on the wave-packet motion.

In the following section, an analysis of the revival structure of
edge-magnetoplasmon wave packets is presented and experimental
detection methods are discussed. 

\section{Dynamics of EMP: the Edge-Magnetoplasmon Wave Packet}
\label{sec3}

The calculations in this section are based on the microscopic theory
of QH samples with sharp edges and $\nu =1/m$. In this case, the edges
have a single branch of chiral bosons. The calculations could readily
be generalized to cases where the microscopic physics of the edge
requires more elaborate boson models, if motivated by future
experimental advances. For a schematic picture of the experimental
geometry we have in mind, see Fig.~\ref{geometry}. We show that
the edge of quantum-Hall systems can be prepared in a state which
is a superposition of EMP narrowly distributed around a peak mode
number. An exact expression for the time evolution of these
wave packets can be obtained and these show an intricate revival
structure. Numerical calculations for a particular set of model
parameters are used to illustrate the latter.

\subsection{Preparation and Evolution of an EMP WP}
\label{sec3A}

We consider a QH sample subject to a time-dependent external 
potential which couples to the charge density of the system. 
We are motivated in large part by an experiment performed by Ashoori
{\it et al.},\cite{ray:prb:92} which used capacitive coupling between
the edge charge density and gates close to the edge of the QH sample
to create EMP excitations and to detect their presence. The
Hamiltonian describing this coupling is
\begin{equation}
H = H_0 + V^{\text{ext}}(t) \quad ,
\end{equation}
with the unperturbed time-independent Hamiltonian for free bosons
(EMP) given by
\begin{equation}\label{freeplasm}
H_0 = \sum_{M>0} \varepsilon^{\text{C}}_{M} \, b_{M}^{\dagger} b_{M}
\quad ,
\end{equation}
and the time-dependent external potential 
\begin{mathletters}\label{potstart}
\begin{eqnarray}
V^{\text{ext}}(t) & = & u(t) \, R \int_{0}^{2\pi} d\theta \,\, 
V^{\text{ext}}(\theta) \, \varrho(\theta)  \\
& = & u(t) \sum_{M>0} (\nu M)^{1/2} \left[ V^{\text{ext}}_{-M} \,
b_{M} + V^{\text{ext}}_{M} \, b_{M}^{\dagger} \right] \,\, .
\label{potend}
\end{eqnarray}
\end{mathletters}
Here, $u(t)$ is the time-dependent voltage pulse applied to the
exciting gate; see Fig.\ \ref{geometry}. The angle $\theta$
parameterizes the coordinate along the edge. The quantity
$\varrho(\theta)$ is the 1D electron density along the edge and
$\varrho_{M}$ is its Fourier transform, while $V^{\text{ext}}(\theta)$
($V^{\text{ext}}_{M}$) models (the Fourier transform of) the coupling
between the gate and the 1D density along the edge, and $\nu$ is the
filling factor. Equation~(\ref{density}) was used to obtain
Eq.~(\ref{potend}). We have in mind the situation where the vertical
separation between the plane containing the exciting gate and the
plane containing the 2D electron layer is much smaller than the
transverse size of the gate so that $V^{\text{ext}}(\theta) \approx 1$
for the portion of the edge under the gate and smoothly falls to zero
outside this region. The time evolution of the system is calculated
most straightforwardly in the Heisenberg picture where the operators
carry all the time-dependence and the states are time-independent: 
\begin{equation}
b^{\pm}_{M}(t) := e^{i H t} \,\, b^{\pm}_{M} \,\, e^{- i H t}.
\end{equation}
(Factors of $\hbar$ are absorbed in our quantum-Hall units.) The
explicit form of these operators may be obtained by solving the
Heisenberg equation of motion. In order to compress the notation,
we write simultaneous equations for creation and annihilation
operators; note that $b^{+}_{M} \equiv b^{\dagger}_{M}$ and
$b^{-}_{M} \equiv b_{M}$. The Heisenberg equation of motion then reads
\begin{mathletters}
\begin{eqnarray}
i \partial_t \,\, b^{\pm}_{M}(t) &=& \left[ \,\, b^{\pm}_{M}(t) \, ,
\, H \,\, \right] \\
&=& \mp \varepsilon^{\text{C}}_{M} \, b^{\pm}_{M}(t) \mp (\nu M)^{1/2}
\, u(t) \, V^{\text{ext}}_{\mp M} \quad ,
\end{eqnarray}
\end{mathletters}
where the second line follows from the first using the commutation
relations for bosonic operators with the Hamiltonian. The solution is 
\begin{equation}\label{pertbos}
b^{\pm}_{M}(t) = \left( b^{\pm}_{M} + B_{\mp M}(t) \right) \,
\exp{\left[ \pm i \varepsilon^{\text{C}}_{M} t \right]} \quad ,
\end{equation}
where $B_{\pm M}(t)$ are complex numbers:
\begin{equation}\label{field}
B_{\pm M}(t) = (\nu M)^{1/2} \, V^{\text{ext}}_{\pm M} \, (\mp i)
\int_{-\infty}^{t} d\tau \,\, u(\tau) \, \exp{\left[ \pm i
\varepsilon^{\text{C}}_{M} \tau \right]}  .
\end{equation}
The undisturbed edge is a collection of independent harmonic
oscillators. When the edge is disturbed by an external potential
which couples to the edge charge density, each harmonic oscillator  
is subject to a different time-dependent external force.
Equation~(\ref{pertbos}) implies that when the edge is initially in
its ground state, the effect of the external force is to put each
oscillator in a coherent state described by the complex field
$B_{\pm M}(t)$.

Inserting Eq.~(\ref{pertbos}) into the expression for the electron
number density along the edge [Eq.~(\ref{density})], we obtain the
Heisenberg-picture expression for the edge density operator:
\widetext
\top{-2.8cm}
\begin{mathletters}
\begin{equation}
\label{rhoexact}
\varrho(\theta, t) = \varrho_0(\theta, t) + \sigma(\theta, t) \quad ,
\end{equation}
where
\begin{eqnarray}
\varrho_0(\theta, t) 
& =& \sum_{M>0} \frac{(\nu M)^{1/2}}{2\pi R} \, \left[ b_{M} \,\,
\exp{\left[ i (M \theta - \varepsilon^{\text{C}}_{M} t) \right] }
\,\, + \,\, b^{\dagger}_{M} \, \, \exp{\left[ -i (M \theta - 
\varepsilon^{\text{C}}_{M} t ) \right] } \right] \quad , \\
\label{sigma} \sigma(\theta, t) &=& \sum_{M>0} \frac{(\nu M)^{1/2}}{2
\pi R} \, \left[ B_{M}(t) \exp{\left[ i (M \theta -
\varepsilon^{\text{C}}_{M} t) \right] } + B_{- M}(t) \exp{\left[ -i
(M \theta - \varepsilon^{\text{C}}_{M} t ) \right] } \right] \quad .
\end{eqnarray}
\end{mathletters}
\bottom{-2.7cm}
\narrowtext
\noindent
Equation~(\ref{rhoexact}) is an {\em exact} expression capturing the
impact of the external time-dependent potential $V^{\text{ext}}(t)$
on the edge charge density. Note that the effect of
$V^{\text{ext}}(t)$ shows up only in Eq.~(\ref{sigma}) for
$\sigma(\theta, t)$, and that $\varrho_0(\theta, t)$ does not
contribute to the expectation value of $\varrho(\theta,t)$ since it
does not conserve boson occupation numbers. As a result the density
response to an external potential is temperature independent.

In the non-equilibrium state created by the excitation pulse, the
time-dependent charge density, given by Eq.~(\ref{rhoexact}) [$\left<
\varrho( \theta, t)\right> = \sigma(\theta, t)$], is identical to the
time-dependent charge density of a linear combination of classical
normal modes. We will henceforth refer to this quantum state of the
edge as an edge-magnetoplasmon wave-packet (EMP WP) state. The
time-evolution of the wave packet is given by Eq.~(\ref{sigma}). As
was discussed above, the spatial structure of the prepared
wave packet as well as its evolution after switching off the potential
is the same at any temperature. 

It is shown in Appendix~\ref{excite} that it is possible to create a
wave packet that has mode numbers strongly peaked around a central
value $\tilde M$. One possible scenario uses a sequence of short
voltage pulses on the gate to excite a superposition of eigenmodes,
in analogy to the use of a short laser pulse to excite a superposition
of Rydberg states. Using this method of excitation, it should be
possible to produce wave packets with mean mode number $\tilde M
\approx 10$ -- $100$.

To detect the wave packet, a second (detecting) gate can be located
at coordinate $\theta_0$ relative to the exciting gate. (Alternately,
the exciting gate could also serve as the detecting gate.) The
charge signal at the detecting gate can be computed as a function of
the delay time $t$: 
\begin{equation}\label{Qt}
Q(t) = R\int_{0}^{2\pi} d\theta\,\, V^{\text{det}}(\theta - \theta_0)
\, \left< \varrho(\theta, t) \right> \quad ,
\end{equation}
where $V^{\text{det}}(\theta) \approx 1 $ under the gate and smoothly
goes to zero outside the gates because of fringe fields. We can
loosely think of $Q(t)$ as the charge under the gate. The voltage
signal on the gate should then be determined by the effective gate
capacitance. (Note that we are not accounting for changes in the
effective interaction between electrons in the 2D ES due to screening
charges on the gates.) In the experiments of
Ref.~\onlinecite{ray:prb:92} the voltage signal is approximately $1
{\rm \mu V}$ per electron under the gate. [If the same gate were used
for excitation and detection we would have $V^{\text{det}}(\theta) =
V^{\text{ext}}(\theta)$.] Inserting Eq.~(\ref{rhoexact}) into
Eq.~(\ref{Qt}), we obtain for $Q(t)$
\begin{equation}
\label{observe}
Q(t) = 2 \Re e \left\{ \frac{1}{2\pi R} \sum_{M>0} Q_{M}(t) \, 
\exp{\left[ i \left( M \theta_0 - \varepsilon^{\text{C}}_{M} t \right)
\right] } \right\}
\end{equation}
with the Fourier components given by 
\widetext
\begin{mathletters}\label{envelope}
\begin{eqnarray}
Q_{M}(t) &=& (\nu M)^{1/2} \, B_{M}(t) \, V^{\text{det}}_{-M} \\ 
&=& \nu M \, V^{\text{ext}}_{M} \, V^{\text{det}}_{-M} \, (- i)
\int_{-\infty}^{t} d\tau \,\, u(\tau) \, \exp{\left[ i
\varepsilon^{\text{C}}_{M} \tau \right]} \,\, . \label{envelope2}
\end{eqnarray}
\end{mathletters}
\bottom{-2.7cm}
\narrowtext
\noindent
This method of creating an EMP WP using a time-dependent gate voltage
and detecting the voltage pulse induced at a second gate by the
time-dependent charge density of the propagating EMP WP is partially
analogous to the phase-sensitive Ramsey method of detection for
Rydberg wave packets.\cite{analogy}

We will refer to the picture of the gate-characteristic functions
$V^{\text{ext}}(\theta)$ and $V^{\text{det}}(\theta)$ explained above,
in which the gate is most sensitive to charges that are located in its
immediate vicinity, as the {\it local-capacitor model}. For the 
calculations reported below we take $V^{\text{ext}}(\theta) =
V^{\text{det}}(\theta) = \exp{[-(\theta \, R / \zeta_{\text{G}})^2]}$
where $\zeta_{\text{G}}$ is the size of the gate. For the Fourier
transforms we find that $V^{\text{ext}}_{M} = V^{\text{det}}_{M} \sim
(\zeta_{\text{G}} / R ) \, \exp{[- (M \, \zeta_{\text{G}} / R )^2]}$.
The factor $V^{\text{ext}}_{M} V^{\text{det}}_{-M}$ in
Eq.~(\ref{envelope2}) then precludes the observation of EMP modes with
wave number $M > R/ \zeta_{\text{G}}$. This fact leads to an important
constraint on the observability of EMP WP. If the excitation scheme
(as discussed in detail in Appendix~\ref{excite}) creates an initial
wave packet that is a superposition of EMP modes with dominant
contribution from the mode with wave number $\tilde M \gg 1$, then
optimal observability of this wave packet using the charge signal
$Q(t)$ requires $\tilde M \zeta_{\text{G}} / R < 1$.
\begin{figure}[hbt]
\epsfxsize=8cm
\centerline{\epsffile{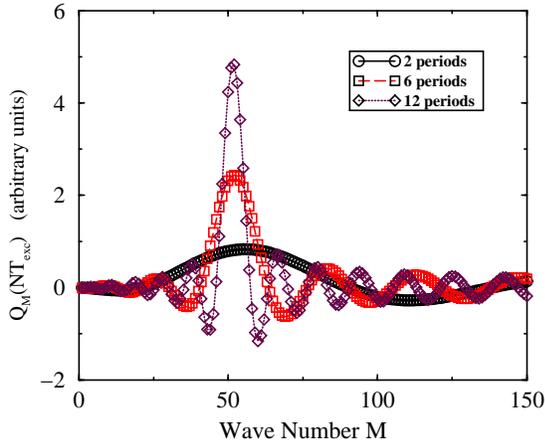}}
\caption{Envelope function $Q_{M}(N T_{\text{exc}})$ for the
wave packet as detected at the end of $N$ excitation pulses of
duration $T_{\text{exc}}$. [See Eqs.~(\ref{observe}) and
(\ref{envelope}) for the definition of the envelope.] The wave packet
is created by multiple short excitation pulses as discussed in 
Appendix~\ref{laser}. These curves were all calculated for
$T_{\text{exc}} = 0.005 \cdot 2\pi R$ and $R = 2500$ in quantum-Hall
units. The figure shows results for different numbers of pulses $N$.
The wave number $\tilde M$ with maximal weight is determined by
$T_{\text{exc}}$. The larger $N$, the sharper the peak in the
envelope function at $M = \tilde M$. For simplicity, it is assumed
that the geometry of the gates does not influence the envelope
function for the range of wave numbers shown.}
\label{lasexc}
\end{figure}

\subsection{Revival Structure} \label{revistruc}

To examine the revival structure of an EMP WP, we first consider the
expectation value for the electron number density along the edge. We
assume that the excitation pulse is turned off at time $\tau$. Then
for $t > \tau$   
\begin{equation}\label{rho}
\left< \varrho(\theta,t) \right> = 2 \Re e \left\{ \sum_{M>0} c_M \, 
e^{iM \theta} e^{-i \varepsilon^{\text{C}}_{M} t} \right\} \quad ,
\end{equation}
where
\begin{equation}\label{cM}
c_M = \frac{\nu M}{2\pi R} \, V^{\text{ext}}_{M} \, \left( -i
\int_{-\infty}^{\tau} d\tau^\prime \,\, u(\tau^\prime) \, \exp{\left[
i \varepsilon^{\text{C}}_{M} \tau^\prime \right]} \right) \quad .
\end{equation}
The coefficients $c_M$ act as weighting functions for the different
modes $M$. Figure~\ref{lasexc} illustrates some possible distributions
for the weightings in $M$ resulting from excitation sequences detailed
in Appendix~\ref{excite}. As a result of the weighting distribution,
only those energies $\varepsilon^{\text{C}}_{M}$ in Eq.~(\ref{rho})
near $\varepsilon^{\text{C}}_{\tilde M}$ will contribute to the sum.
This permits a Taylor expansion of $\varepsilon^{\text{C}}_{M}$ around
the central value $\varepsilon^{\text{C}}_{\tilde M}$.

The derivative terms in this expansion define the time scales that
control the evolution and revival structure of the wave packet, as
discussed above in the section on Rydberg wave packets. In this case 
the expressions for the time scales are [cf.\ Eqs.~(\ref{gentimes})]:
\begin{mathletters}
\label{times}
\begin{eqnarray}
T_{\text{cl}} &=&  \frac {2 \pi} {\vert (
\varepsilon^{\text{C}}_{\tilde M})^\prime \vert} = \frac {2 \pi R}
{(\ln \frac R {\alpha \tilde M} - 1)} \quad , \\
t_{\text{rev}} &=&  \frac {2 \pi} {\frac 1 2 \vert ( 
\varepsilon^{\text{C}}_{\tilde M})^{\prime\prime} \vert} =  4 \pi R
\tilde M \quad , \\  t_{\text{sr}} &=& \frac {2 \pi} {\frac 1 6 \vert 
(\varepsilon^{\text{C}}_{\tilde M})^{\prime\prime\prime} \vert} =
12 \pi R \tilde M^2 \quad .
\end{eqnarray}
\end{mathletters}
The primes denote derivatives with respect to $M$.
Equations~(\ref{times}) define the classical orbit period, the revival
time, and the superrevival time, respectively, for EMP WP.

For large values of $\tilde M \approx 10$ -- $100$ and typical values
of $R \approx 10^4$, we find $t_{\text{sr}} \gg t_{\text{rev}} \gg
T_{\text{cl}} \gg T_{\text{ph}} := 2\pi/
\varepsilon^{\text{C}}_{\tilde M}$.
Therefore, for times $t \ll t_{\text{rev}}$, we can approximate
$\left< \varrho(\theta,t) \right> \approx 2 \Re e \left\{ {\tilde
\varrho}_{\text{cl}} (\theta,t) \right\}$, where
\begin{eqnarray}
\tilde \varrho_{\text{cl}}(\theta,t) &=&  e^{i (\tilde M \theta - 2
\pi t / T_{\text{ph}} )} \, \sum_{M>0} c_M \, e^{i (M-\tilde M) (
\theta - 2 \pi t / T_{\text{cl}} ) } \nonumber \\
&=& e^{i (\tilde M \theta - 2 \pi t / T_{\text{ph}} )} \cdot
\varrho_{\text{cl}}(\theta,t) \quad.
\end{eqnarray}
Here $T_{\text{ph}} = 2 \pi R / (\tilde M \, \ln[R/(\alpha \tilde M)
])$. Thus $\tilde \varrho_{\text{cl}}(\theta,t)$ is the product of a
rapidly oscillating\cite{wiggles} (both in space and time) phase
factor and a more slowly varying periodic envelope function. The
period of the rapid spatial oscillations is $2\pi/\tilde M$, whereas
$T_{\text{ph}}$ is the period of the rapid temporal oscillations. If
the additional phase factor were not present, the charge density in
this approximation would circulate around the edge without distortion
and with period $T_{\text{cl}}$. This motion resembles the classical
drift motion\cite{jackson} of the cyclotron orbit of a charged
particle in a strong magnetic field; it is perpendicular to both
magnetic and electric fields and has speed $v_{\text{dr}} = c E / B$.
In the present case,the electric field is perpendicular to the edge
of the 2D ES so that the drift is along the edge and the classical
period is\cite{classcaveat} $T_{\text{cl}} \approx 2 \pi R /
v_{\text{dr}}$. The electric field which yields our classical orbit
period is that from the neutralizing background required to stabilize
a macroscopic system of charged particles.\cite{classcaveat} Because
of the additional phase factor this classical motion appears as the
envelope of a more rapid oscillation of edge charge density. In what
follows, we focus on the evolution of this classical envelope function
at longer times. 

For times greater than $T_{\text{cl}}$, the second-derivative term
will eventually contribute to the time-dependent phase, and we expect
a revival structure analogous to the fractional and full revivals of
Rydberg wave packets. Indeed, at the times $t = m\, t_{\text{rev}}/n$,
with $m$ and $n$ relatively prime ($m\le n$), we can write the sum
over $M$ [in Eq.~(\ref{rho})] as a sum of subsidiary functions,
\begin{equation}
\label{rhofrac}
\left< \varrho(\theta,t) \right> \approx 2 \Re e \left\{ e^{i (\tilde
M \theta - 2 \pi t / T_{\text{ph}} )} \,\, \sum_{s=0}^{l-1} a_s \,
\varrho_{\text{cl}}(\theta,t + \frac s l T_{\text{cl}}) \right\}
\, .
\end{equation}
The coefficients $a_s$ are given by
\begin{equation}
\label{as}
a_s = \frac 1 l \sum_{M^\prime = 0}^{l-1} \exp[2 \pi i \frac m n
(M^\prime)^2] \, \exp [2 \pi i M^\prime \frac s l ] \quad ,
\end{equation}
with
\begin{equation}
\label{lcases}
l =\cases{\frac n 2&if~~$n = 0~~ ({\rm mod}~4)~~$,\cr
        n&if~~$n \ne 0~~ ({\rm mod}~4)~~$.\cr }
\end{equation}
The moduli $\vert a_s \vert$ are equal for all $l$. Therefore, the
sum over $s$ in Eq.~(\ref{rhofrac}) consists of an equally weighted
sum of subsidiary functions that are shifted in time by fractions of
the classical period $T_{\text{cl}}$. This demonstrates that at the
fractional revivals the EMP WP can be written as a sum of subsidiary
waves. The motion of the wave packet is periodic with a period equal
to a fraction of $T_{\text{cl}}$. The full revival occurs when
$m=n=1$. In this case $l=1$, $a_0 = 1$ is the only nonzero coefficient
in Eq.~(\ref{rhofrac}), and the wave-packet sum consists of a single
term which has the same form as the initial wave packet. This analysis
for $t \leq t_{\text{rev}}$ ignores contributions from the
higher-order terms in the phase. As a result of these higher-order
terms, the full revival is not a perfect revival in that it does not
exactly equal the initial wave packet. Including the third-order term
in the time-dependent phase, which depends on the time scale
$t_{\text{sr}}$, would lead to an analysis of the superrevival
structure analogous to that performed previously for Rydberg
wave packets.

The experimentally measured quantity is not $\varrho (\theta,t)$ but
the charge signal $Q(t)$ at the detecting gate defined in
Eq.~(\ref{observe}). In this expression the weighting coefficients
$Q_M(t)$ are constant after the excitation pulse has been turned
off. Expanding $\varepsilon^{\text{C}}_{M}$ in a Taylor series in $M$
around the value $\tilde M$ shows that $Q(t)$ has a time dependence
similar to $\left< \varrho (\theta,t) \right>$. The full and
fractional periodicities in $T_{\text{cl}}$ should therefore be
exhibited in the time dependence of the charge signal $Q(t)$.

\subsection{Example}
As an example, we consider an EMP WP with $\tilde M = 50$. In the
next section, we discuss the experimental feasibility of creating and
observing a wave packet with mode numbers in this range. In realistic
samples, such a wave packet will be damped and lose phase coherence
due to electron-phonon interactions, and possibly also due to
interactions with electrons in the bulk of the two-dimensional system
when it has localized low-energy excitations. In this section, we
ignore for the moment damping and loss of phase coherence of the
wave packet and examine the resulting idealized revival structure for
times up to $t_{\text{rev}}$.

As shown in Appendix~\ref{excite}, there are many possible weighting
distributions for the coefficients $c_M$, depending on the
experimental configuration. The analysis of the revivals given above
requires only that the coefficients $c_M$ be strongly peaked around
a central value $\tilde M$. For the sake of definiteness we take a 
particular example in this subsection, modelling the coefficients
$c_M$ by a Gaussian distribution centered on $\tilde M = 50$ with
width $\sigma_M = 2$. We also set $\alpha = 1$ and $R = 10^4$. From
Table~\ref{table_nsu} for typical magnetic field strengths this
corresponds to a sample with linear dimension $\sim 100$~$\rm \mu m$
and time scales starting from the nanosecond range. To be specific,
we use $\nu = 1$ and $B = 10$~T, which fixes our time scales to
$T_{\text{cl}} \approx 2.2$~ns, $t_{\text{rev}} \approx 942$~ns, and
$t_{\text{sr}} \approx 141$~$\rm \mu s$. In the chiral-boson model the
time evolution is independent of the overall strength of the signal
so we plot results for $\left< \varrho(\theta,t) \right>$ calculated
from Eq.~(\ref{rho}) as a function of the angle $\theta$ at various
times in arbitrary units. 

\begin{figure}[hbt]
\epsfxsize=16cm
\centerline{\epsffile{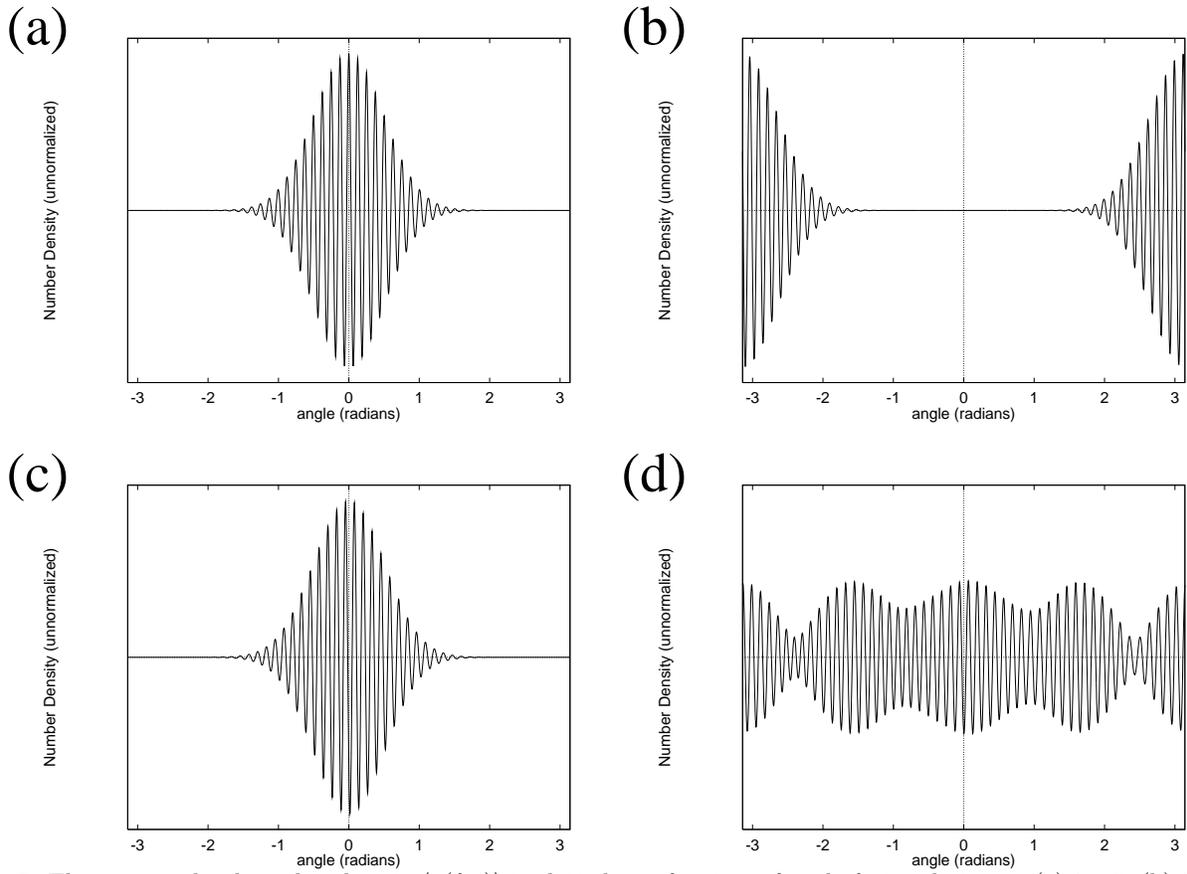}}
\caption{The unnormalized number density $\left< \rho (\theta,t)
\right>$ is plotted as a function of angle $\theta$ at early times.
(a)~$t = 0$, (b)~$t = T_{\text{cl}}/2$, (c)~$t = T_{\text{cl}}$, and
(d)~$t = 50 \, T_{\text{cl}}$.}
\label{rhoperiodic}
\end{figure}

\begin{figure}[hbt]
\epsfxsize=18cm
\centerline{\epsffile{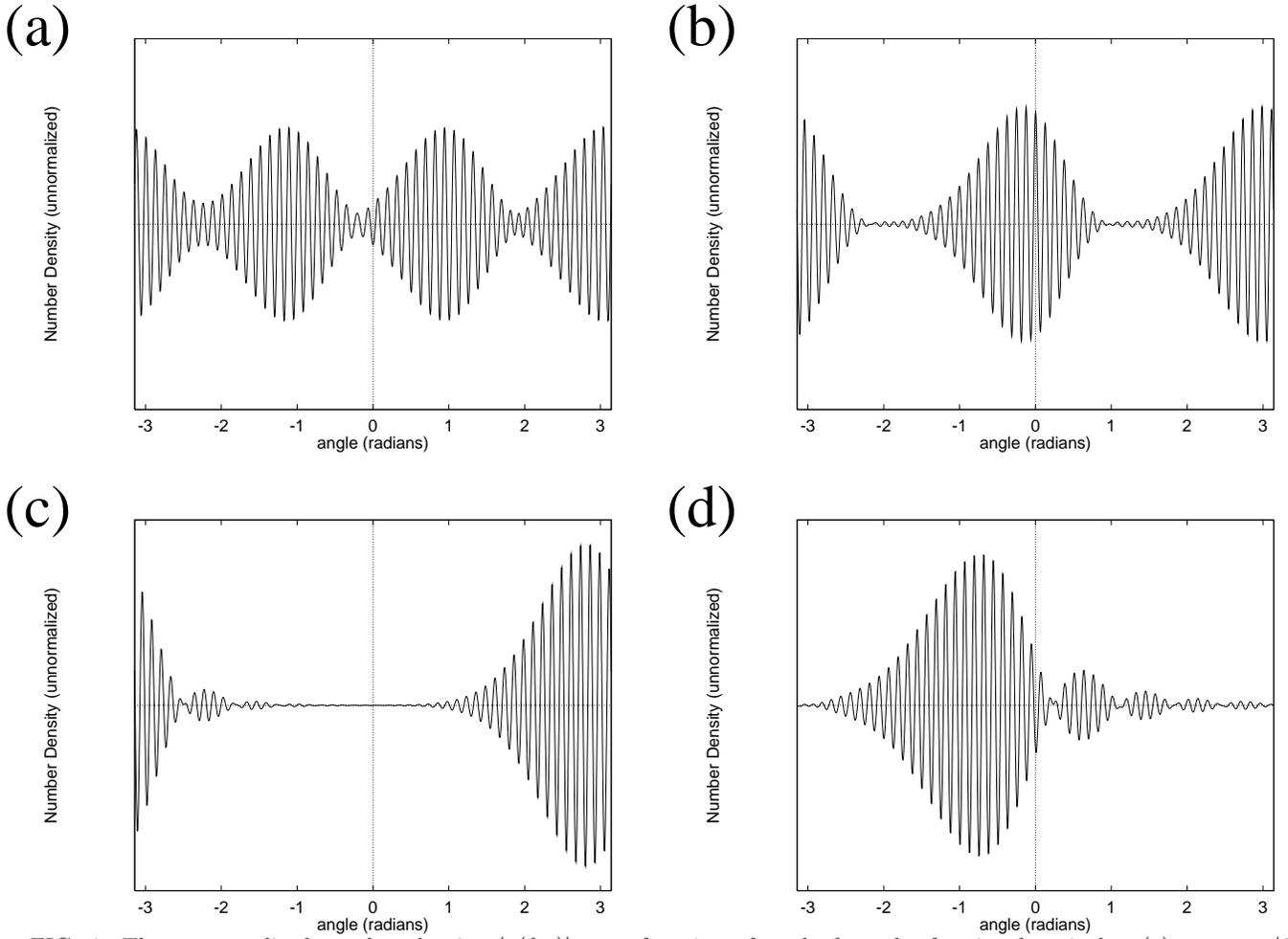}}
\caption{The unnormalized number density $\left< \rho (\theta,t)
\right>$ as a function of angle $\theta$ at the fractional revivals.
(a)~$t = t_{\text{rev}}/6$, (b)~$t = t_{\text{rev}}/4$,
(c)~$t = t_{\text{rev}}/2$, and (d)~$t = t_{\text{rev}}$.}
\label{rhorevivals}
\end{figure}

Figure~\ref{rhoperiodic}a shows the initial wave packet for the
Gaussian model considered in this section, which sums to a smooth
envelope on an oscillating background. Figures~\ref{rhoperiodic}b and
\ref{rhoperiodic}c show the wave packet at times $T_{\text{cl}}/2$
and $T_{\text{cl}}$, respectively. It is seen that initially the
wave packet maintains its shape and exhibits the classical
periodicity. However, for times beyond $T_{\text{cl}}$
quantum interference effects commence. Figure~\ref{rhoperiodic}d shows
the wave packet at $50 \, T_{\text{cl}}$. After 50 classical periods,
the wave packet has spread and is no longer localized.
\widetext
\narrowtext
\noindent
Figure~\ref{rhorevivals} shows the fractional and full revivals. At
$t = t_{\text{rev}}/6$, there are three nonzero terms in the sum in
Eq.~(\ref{rhofrac}). The wave packet splits into three equally
weighted subsidiary waves as illustrated in Fig.~\ref{rhorevivals}a.
The motion of the wave packet is periodic with period $T_{\text{cl}}/
3$. Figure~\ref{rhorevivals}b shows the wave packet at
$t_{\text{rev}}/4$, at which time it consists of a sum of two distinct
subsidiary waves. Here, the motion is periodic with period
$T_{\text{cl}}/2$. A full revival first occurs at $t_{\text{rev}}/2$.
Figure~\ref{rhorevivals}c shows that a single wave packet has formed.
This revival has the quantum-mechanical characteristic of being
one-half cycle out of phase with the classical motion.\cite{ap}
Only at the time $t_{\text{rev}}$ does a single wave packet form that
is in step with the corresponding classical periodic motion. The full
revival at $t_{\text{rev}}$ is shown in Fig.~\ref{rhorevivals}d. It
evidently does not exactly match the initial wave packet. A smaller
subsidiary wave packet has also formed.

The revival structure as a function of time can also be observed by
computing the charge signal $Q(t)$. This is the observable that would
be measured in an experiment. We evaluate Eq.~(\ref{observe}), using
the same Gaussian weighting as above and ignoring the overall
normalization.

\begin{figure}[hbt]
\epsfxsize=18cm
\centerline{\epsffile{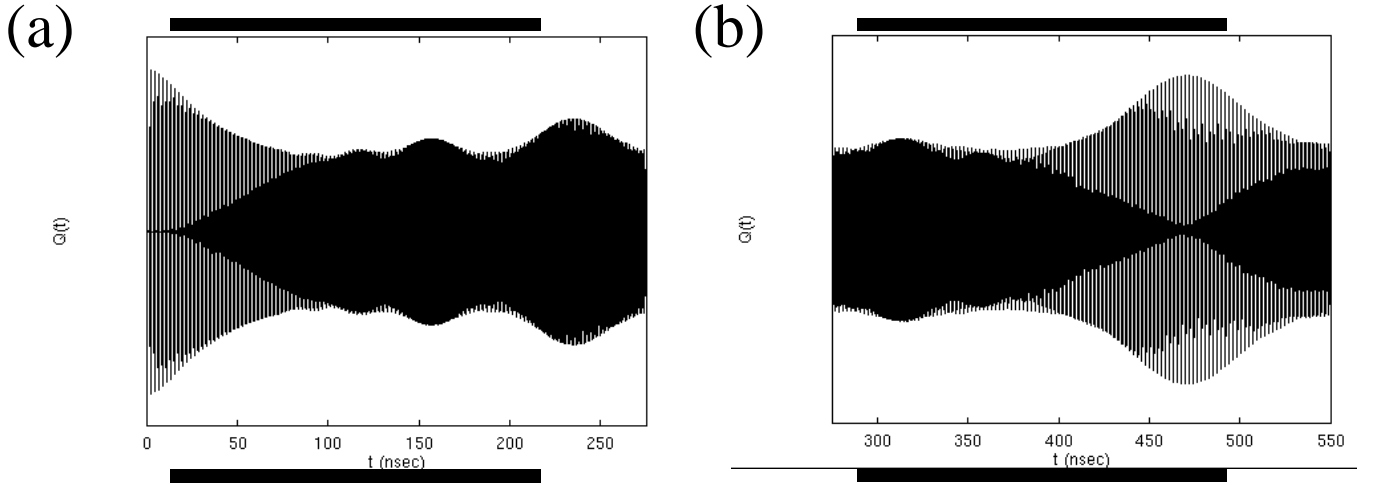}}
\vspace{3mm}
\caption{The charge signal $Q(t)$ is plotted as a function of time in
nanoseconds. Figures~(a) and (b) illustrate the large-scale
periodicities in $Q(t)$ up to times of order $t_{\text{rev}}/2
\approx 471$~ns. The rapidly oscillating part of the signal is not
resolved in these figures and the individual peaks at short times are
separated by the classical period. At partial revivals the period of
the revived wave packet is shorter than the classical period. These
shorter periods are not resolved in this figure resulting in nearly
solid portions of the figure. In this case the extent of the solid
region reflects the amplitude of the revived wave packet.} 
\label{Qbig}
\end{figure}

Figures~\ref{Qbig}a and \ref{Qbig}b show $Q(t)$ as a function of time
for times up to and just beyond $t_{\text{rev}}/2$. Here, the revival
structure is revealed through the periodicity of $Q(t)$. In
Fig.~\ref{Qbig}a, the initial motion is clearly periodic with period
$T_{\text{cl}} \approx 2.2$~ns. However, as the wave packet spreads
and collapses, the peaks in $Q(t)$ disappear and fractional revivals
start to emerge. The peaks at $t_{\text{rev}}/4 \approx 235$~ns have
half the amplitude and periodicity of the initial peaks. This is
because the wave packet has reformed into two distinct parts. The
peaks at $t_{\text{rev}}/6 \approx 157$~ns have period
$T_{\text{cl}}/3$ and even smaller amplitude, corresponding to the
formation of three subsidiary wave packets. Figure~\ref{Qbig}b shows
the full revival at $t_{\text{rev}}/2 \approx 471$~ns, which is
one-half cycle out of phase with the classical motion. Here, the
amplitude of the peaks matches that of the initial peaks, and the
period is again $T_{\text{cl}}$.

\begin{figure}[hbt]
\epsfxsize=18cm
\centerline{\epsffile{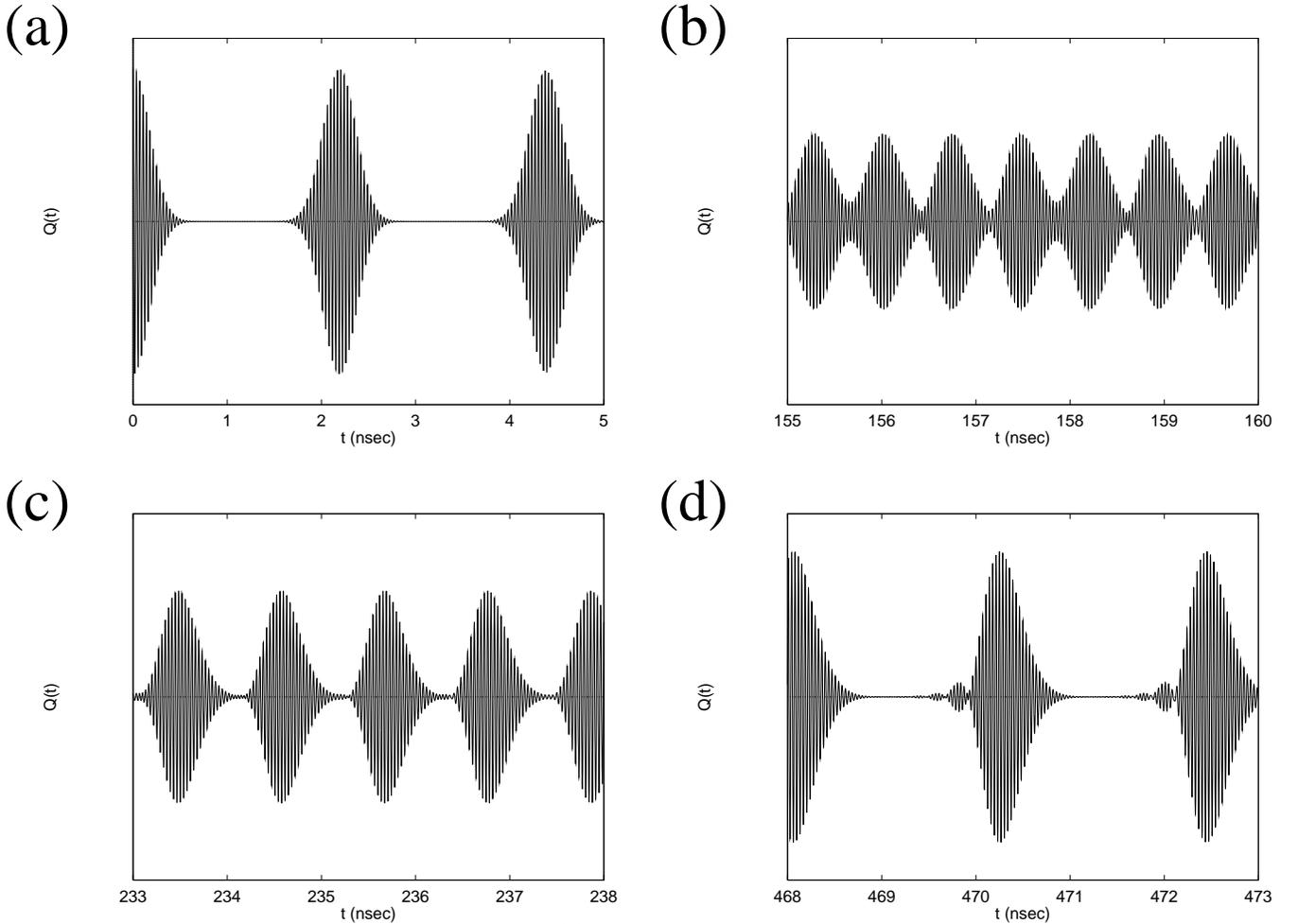}}
\caption{An enlargement of Fig.~\ref{Qbig} at various times. (a)~the
initial classical cycles for $0 \le t \le 5$~ns, revealing the
presence of the rapid oscillations in $Q(t)$. Figures~(b), (c), and
(d) show enlargements of Fig.~\ref{Qbig} near the times $t =
t_{\text{rev}}/6$, $t = t_{\text{rev}}/4$, and $t = t_{\text{rev}}/2$,
respectively, corresponding to the formation of 3, 2, and 1
subsidiary waves at the fractional revivals.}
\label{Qenlarged}
\end{figure}

Figure~\ref{Qenlarged} enlarges some of the regions in
Fig.~\ref{Qbig}. Figure~\ref{Qenlarged}a shows the first few
classical cycles for times up to 5~ns. Evidently, the individual
peaks with period $T_{\text{cl}}$ actually consist of envelopes of
highly oscillatory signals.\cite{wiggles,analogy} These rapid
oscillations cannot be resolved in Fig.~\ref{Qbig}. The period of the
rapid oscillations is $T_{\text{ph}} \approx 0.036$~ns.
Figure~\ref{Qenlarged}b shows an enlargement of $Q(t)$ near
$t_{\text{rev}}/6$. The period of the envelope peaks is
$T_{\text{cl}}/3 \approx 0.73$~ns. Figure~\ref{Qenlarged}c shows the
charge signal near $t_{\text{rev}}/4$, where two wave packets are
moving with period $T_{\text{cl}}/2 \approx 1.1$~ns. The charge signal
near $t_{\text{rev}}/2$ is shown in Fig.~\ref{Qenlarged}d. The period
is again $T_{\text{cl}}$, but some distortion of the signal is
evident in comparison to the initial peaks in Fig.~\ref{Qenlarged}a.
\widetext
\narrowtext
\noindent

This example demonstrates explicitly the formation of full and
fractional revivals of EMP WP. The generic features are the same as
for Rydberg wave packets. At the fractional revivals, the EMP WP is
equal to a sum of subsidiary wave packets that move with a
periodicity equal to a fraction of $T_{\text{cl}}$. The semiclassical
behavior as well as the revivial structure of EMP WP can be detected
experimentally by measuring periodicities in the envelope that
modulates rapid oscillations in the charge signal $Q(t)$. These
oscillations are the analog of the Ramsey fringes in
Rydberg-wave-packet experiments which use the phase-sensitive method
of detection.

\subsection{Fluctuations: quantum and thermal}

The EMP WP is a many-mode coherent state: each plasmon mode represents
a one-dimensional quantum harmonic oscillator, and the creation scheme
described above results in each oscillator mode being excited into a
coherent state. The above-mentioned detection scheme yields an average
charge signal $Q(t)$ which reflects purely classical behavior of the
EMP modes. Quantum fluctuation effects appear only in the noise
spectrum of the charge signal. 

In our detection scheme, we measure the expectation value of the
charge operator $\hat Q(t)$:
\begin{mathletters}
\begin{eqnarray}
\hat Q(t) &=& R \int_{o}^{2\pi} d\theta \, V^{\text{det}} (\theta -
\theta_0) \, \varrho(\theta, t) \\
&=& \sum_{M>0} \left[ \nu M \right]^{1/2} \, \left\{
V^{\text{det}}_{-M} \, b_{M}(t) \, e^{i M \theta_0} + \rm h.c.
\right\} \,\,\, .
\end{eqnarray}
\end{mathletters}
It is possible to rewrite $\hat Q(t)$ as the sum of a term [$=: \delta
\hat Q(t)$] that has vanishing expectation value and contributes only
to fluctuations, and a term [$\equiv Q(t)$] which does not fluctuate
at all and equals the expectation value of the charge operator: $\hat
Q(t) = \delta \hat Q(t) + Q(t)$. Explicitly, we find
\begin{equation}
\delta \hat Q(t) = \sum_{M>0} \left[ \nu M \right]^{1/2} \, \left\{
V^{\text{det}}_{-M} \, b_{M} \, e^{i [ M \theta_0 -
\varepsilon^{\text{C}}_{M} t ]} + \rm h.c.  \right\} \quad ,
\end{equation}
and $Q(t)$ was defined in Eq.~(\ref{observe}). The variance of the
charge signal is readily evaluated:
\begin{mathletters}\label{noise}
\begin{eqnarray}
\big< \big[ \Delta \hat Q(t) \big]^2 \big> &=& \big< \big[ \Delta
\hat Q(t) \big]^2 \big>_{\text{qu}} + \big< \big[ \Delta \hat Q(t)
\big]^2 \big>_{\text{th}} \,\, , \\ \label{zeronoise}
\big< \big[ \Delta \hat Q(t) \big]^2 \big>_{\text{qu}} &=&
\sum_{M>0} \nu M \, | V^{\text{det}}_{M} |^2 \quad , \\
\big< \big[ \Delta \hat Q(t) \big]^2 \big>_{\text{th}} &=&
2 \sum_{M>0} \nu M \, | V^{\text{det}}_{M} |^2 \, n_{M}^{(0)}
\quad , \label{thermnoise}
\end{eqnarray}
\end{mathletters}
with $n_{M}^{(0)}$ being the thermal-equilibrium occupation number of
the plasmon mode labelled by quantum number $M$.
Equations~(\ref{zeronoise}) and (\ref{thermnoise}) represent the noise
due to quantum and thermal fluctuations, respectively.

We assume the detector characteristics to be determined primarily by
geometrical properties, such as the dimension $\zeta_{\text{G}}$ of
the capacitor plate. Within the local-capacitor model, we have
$V^{\text{det}}_{M} \sim (\zeta_{\text{G}}/ R) \, \exp{
\{ - [M \, \zeta_{\text{G}} / R ] ^2 \} }$ which yields the result
\begin{equation}
\big< \big[ \Delta \hat Q(t) \big]^2 \big> \approx \nu \, \times \,\,
\left\{ \begin{tabular}{cl}
1 & if $\zeta_{\text{T}} > \zeta_{\text{G}}$ \\
$\zeta_{\text{G}}/ \zeta_{\text{T}}$ & otherwise
\end{tabular}
\right.
\end{equation}
where the thermal length $\zeta_{\text{T}} := v_{\text{dr}} / (
\sqrt{4\pi} k_B T)$. The regime where $\zeta_{\text{T}} >
\zeta_{\text{G}}$ is dominated by quantum fluctuations, whereas
thermal fluctuations are more important if $\zeta_{\text{T}} <
\zeta_{\text{G}}$.

In order to judge the importance of the quantum and thermal
fluctuations for experiments detecting the semiclassical behavior
and revival structure of EMP WP, we compare the magnitude of the
fluctuations to the amplitude of the charge signal $Q(t)$. As an
order-of-magnitude estimate, we find for the case of an EMP WP which
was created using the multiple-short-pulse technique (see
Appendix~\ref{laser}):
\begin{equation}
|Q(t)|^{\text{max}} \sim \nu \,\, \frac{\tilde M \,
\zeta_{\text{G}}}{R} \,\, \frac{u_0 \,
\zeta_{\text{G}}}{v_{\text{dr}}} \quad .
\end{equation}
Typical drift velocities $v_{\text{dr}}$ are of the order of $5 \times
10^5$~m/s, so that we get a numerical estimate
\begin{equation}
|Q(t)|^{\text{max}} \sim 3 \, u_0[\rm mV] \, \zeta_{\text{G}}[\mu m]
\,\, \times \,\, \nu \,\, \frac{\tilde M \,\zeta_{\text{G}}}{R}
\quad .
\end{equation}
Here, $u_0$ denotes the amplitude of the voltage pulse which created
the EMP WP. Remember that $\tilde M \zeta_{\text{G}} / R < 1$ is
required to enable the detection of the EMP WP using the
charge signal. For a signal-to-noise ratio greater than unity, the
amplitude of the voltage pulse has to satisfy
\begin{equation}
u_0 \gg \frac{R}{\tilde M \zeta_{\text{G}}} \,\,
\frac{v_{\text{dr}}}{\zeta_{\text{G}}} \,\,\, \rm max \left\{ 1, (
\zeta_{\text{G}} / \zeta_{\text{T}} )^{1/2} \right\}
\quad ,
\end{equation}
or based on the numerical estimate above
\begin{equation}
u_0[\rm mV] \gg \frac{0.3}{\zeta_{\text{G}}[\mu m]} \,\,\times \,\,
\frac{R}{\tilde M \zeta_{\text{G}}} \quad .
\end{equation}

\section{Finite Lifetime of Edge-Magnetoplasmon Wave Packets}
\label{sec4}

The previous section analyzed the formation and revival structure of
EMP WP, assuming an infinite life time for the EMP which form the
wave packet. It was shown that thermal effects have no influence on
the preparation and evolution of these wave packets apart from
contributing to fluctuations. The revival structures we have
discussed require coherent evolution of the EMP system. In realistic
systems, electrons in the 2D ES will be coupled to the semiconductor
host material via various physical processes. For experimentally
realistic temperatures and parameter ranges in semiconductors, the
most important process will typically be the coupling of electrons to
3D acoustic phonons. This coupling leads to a finite life time for
EMP, which we calculate in this section. Comparing the life time to
the relevant time scales for semiclassical behavior ($T_{\text{cl}}$)
and for revivals ($t_{\text{rev}}$), we can determine the
observability of these effects.

\subsection{Plasmon-Phonon Coupling -- General}

The electron-phonon interaction is specified by the following
contribution to the Hamiltonian,\cite{mahan}
\begin{equation}\label{elecphon}
H^{\text{el-ph}} = \sum_{\vec q, \lambda} M^{\lambda}(\vec q) \left(
a^{\dagger}_{- \vec q, \lambda} + a_{\vec q, \lambda} \right)
\varrho^{\text{3D}}_{\vec q} \quad ,
\end{equation}
where the operators $a^{\dagger}_{\vec q, \lambda}$
($a_{\vec q, \lambda}$) create (annihilate) phonons with 3D
wave vector $\vec q$, polarization label $\lambda$, and normal-mode
frequency $\omega^{\lambda}_{\vec q}$. Here,
$\varrho^{\text{3D}}_{\vec q}$ denotes the 3D electron density
in the QH sample and $M^{\lambda}(\vec q)$ is a coupling constant 
whose numerical value is known in most materials of interest. 
Since EMP are collective fluctuations of the electron density at
the edge of our sample, the electron-phonon interaction leads to an
effective coupling between phonons and EMP. By identifying the
contribution to the 3D electron density from EMP we are able to
derive a Hamiltonian which describes the coupling between the EMP
and phonon systems. For example, for the disk geometry described in
Fig.~\ref{geometry}, the Hamiltonian has the form
\begin{table}[hbt]
\setlength{\parindent}{0.0in}
\caption{Parameters of GaAs/\protect{Al$_x$Ga$_{1-x}$As}
heterostructures, according to
Ref.~\protect\onlinecite{lyo:prb:88}.}
\begin{tabular}{rcl}
Mass density & $\rho$ & $5300\,\rm kg / m^{3}$ \\
Longitudinal sound velocity & $c^l$ & $5140\,\rm m/s$ \\
Transversal sound velocity & $c^t$ & $3040\,\rm m/s$ \\
Deformation potential & $D$ & 9.3 eV \\
Piezoelectric constant & $h_{\text{14}}$ & $1.2\times 10^9\,\rm V/ m$
\end{tabular} 
\label{numbers}
\end{table}
\widetext
\begin{equation}\label{coupli}
H^{\text{pl-ph}} = \sum_{M>0 \atop \vec q, \lambda} 
C^{\lambda}_{M}(\vec q) \left( a_{\vec q, \lambda} +
a^{\dagger}_{- \vec q, \lambda} \right) \left( b_{M} \, e^{i M 
\theta_{\vec q}} + b^{\dagger}_{M} \, e^{- i M \theta_{\vec q}}
\right) \quad ,
\end{equation}
\bottom{-2.7cm}
\narrowtext
\noindent
where $\theta_{\vec q}$ denotes the azimuthal angle of the wave vector
$\vec q$ in the plane of the 2D ES. Details of the derivation of
Eq.~(\ref{coupli}) and an analytic expression for the coupling
coefficients $C^{\lambda}_{M}(\vec q)$ are given in
Appendix~\ref{plasphon}. It is possible to derive an expression 
similar to Eq.~(\ref{coupli}) which is valid for the strip geometry; 
this case is also discussed in Appendix~\ref{plasphon}.

To investigate the effect of plasmon-phonon coupling on the evolution
of an EMP WP, we consider a single-plasmon Matsubara Green's function defined by  
\begin{equation}\label{greensfunc}
{\cal G}_{M}(\tau) = - \big< \mbox{T}_\tau \, b_{M}(\tau)
b_{M}^{\dagger} (0) \big> \quad ,
\end{equation}
and the Fourier transform ${\cal G}_{M}(\omega)$ of the retarded
Green's function, which is obtained from the Fourier transform of 
${\cal G}_{M}(\tau)$ by continuation to real frequencies. In the
absence of EMP-phonon coupling, ${\cal G}_{M}(\omega)$ reduces to the
well-known result\cite{mahan} for free bosons: ${\cal G}_{M}^{0}(
\omega) = [\omega - \varepsilon_{M}^{\text{C}} + i \delta ]^{-1}$
reflecting the fact that EMP are the well-defined excitations of the
system. The presence of phonons causes damping (and an energy shift)
of the EMP, leading to the modified result ${\cal G}_{M}(\omega) =
[\omega - \tilde \varepsilon_{M} + i \, \Gamma_{M}/2]^{-1}$ with
$\Gamma_{M}^{-1}$ being the life time of the EMP with wave number $M$.
We find
\begin{equation}\label{damprate}
\Gamma_{M} = 2 \pi \sum_{\vec q, \lambda} \left| C^{\lambda}_{M}
(\vec q) \right|^2 \,\, \delta \big( \varepsilon_{M}^{\text{C}} -
\omega_{\vec q}^{\lambda} \big) \quad .
\end{equation}
See Appendix~\ref{selfen} for details of the calculation.

The meaningful quantity to assess the effect of EMP-phonon coupling
on the propagation of EMP WP is $\Gamma_{\tilde M}^{-1}$, which we
call the life time of the EMP WP.

At energy and wave-vector scales appropriate for the observation of
EMP WP, acoustic phonons with dispersion $\omega^{\lambda}_{\vec q}
= q\, c^{\lambda}$ are most important. In a polar semiconductor, both
scattering from a deformation potential and piezoelectric effects
contribute to the electron-phonon coupling, whereas in a non-polar
semiconductor the piezoelectric part is absent. In the following two
sub-sections, we discuss these two cases separately.

\subsection{Polar Semiconductors: GaAs}

At the long wavelengths used to construct EMP WP, piezoelectric
coupling dominates the deformation-potential scattering of electrons.
We therefore neglect the contribution from the deformation potential
in this sub-section. Evaluation of Eq.~(\ref{damprate}) for the disk
case yields 
\begin{equation}
\Gamma_{M} \approx \left( \frac{h_{14}}{\tilde c} \right)^2 
\frac{\epsilon}{\rho} \, \frac{M}{R} \quad ,
\end{equation}
with $\epsilon$ being the semiconductor bulk dielectric constant, and
the piezoelectric coupling constant denoted by $h_{14}$. Here, the
quantity $\tilde c$ is of the order of the speed of sound. This
estimate is based on the observation that the drift velocity of the
EMP WP is typically $\sim 100$ times larger than the speed of sound
so that typical projections of the phonon wave vector onto the 2D
plane lead to large Bessel-function arguments. (Typical phonon
wavelengths are very small compared to the size of the disk.) Note
that modes with higher wave number decay faster. 

The relevant quantities to examine when assessing the possibility to
observe wave-packet revivals are: 
\begin{mathletters}\label{phonlimit}
\begin{eqnarray}
\frac{\Gamma_{\tilde M}^{-1}}{T_{\text{cl}}} &\sim& \left(
\frac{\tilde c}{h_{\text{14}}} \right)^2 \frac{\rho}{\epsilon} \,\,
\frac{\ln [R/\tilde M]}{\tilde M} \sim 100 \,\, \frac{\ln [R / \tilde
M]}{\tilde M} \,\,\, , \\ \label{revobs}
\frac{\Gamma_{\tilde M}^{-1}}{t_{\text{rev}}} &\sim& \left(
\frac{\tilde c}{h_{\text{14}}} \right)^2 \frac{\rho}{\epsilon} \,\,
\frac{1}{\tilde M^2} \sim \left(\frac{7}{\tilde M} \right)^2 \quad .
\end{eqnarray}
\end{mathletters}
Equations~(\ref{phonlimit}) give respectively the number of classical
periods and the number of revivals which occur in the mean free time
of an EMP WP. We see that, unless $\tilde M$ is very large, the
initial periodicity should be observable. However, the likelihood of
seeing revivals appears to be quite remote. Equation~(\ref{revobs})
shows that the wave packet loses coherence before it can revive. The
typical values of parameters for GaAlAs/GaAs-heterostructures used in
the calculation of the EMP WP life time are taken from
Ref.~\onlinecite{lyo:prb:88} and are shown in Table~\ref{numbers}.

\subsection{Interpretation of Previous Experiment} 

It is interesting to reexamine the experiments described in 
Ref.~\onlinecite{ray:prb:92} in the light of these expressions. In
that work measurements were made on a QH sample in GaAs with the
geometry sketched in Fig.~\ref{geometry} at temperature $T=0.3$~K.
The values of the relevant parameters were: filling factor $\nu=1$,
magnetic-field strength $B=5.1$~T ({\it i.e.} magnetic length $\ell
\approx 11$~nm), $R = 270\,\rm \mu m \approx 2.4 \times 10^4 \, \ell$
and $\zeta_{\text{G}} \sim 10\,\rm \mu m$. A single voltage pulse with
amplitude $u_0 = 50$~mV and duration $T_{\text{exc}} = 100$~ps was
applied to create the initial wave packet. The latter was observed to
move around the disk sample with period $T_{\text{cl}} \approx 4$~ns
while spreading rapidly. (Fewer than ten cycles can be discerned
before the signal vanishes in the noise.) The data do {\em not} seem
to consist of a rapid oscillation that is modulated by an envelope.

We believe that the EMP WP excited in this experiment was composed 
primarily of modes with $M < 5$, with $M=1$ possibly having the
largest amplitude. Our analysis of the time scales would give
$T_{\text{ph}}\approx T_{\text{cl}} \approx t_{\text{rev}} / 20$. In
our interpretation of this experiment, the period of the rapid
oscillation in the charge signal and the classical period (which
determines the periodicity of the {\em envelope function} that
modulates the rapidly oscillating charge signal) are nearly equal, and
the revival time is just one order of magnitude larger. The absence of
a rapid oscillation in the charge signal results from the
near-equality of $T_{\text{ph}}$ and $T_{\text{cl}}$. The decay of the
charge signal in $\sim 10$ classical periods can be consistently
explained as being due to the spreading of the wave packet due to the
non-linear dispersion of EMP. However, because of the small value of
$\tilde M$ in this experiment, interference between first- and
higher-order terms in the expansion of the non-linear dispersion of
the EMP around $\tilde M$ would be expected to (and apparently does) 
obscure the fractional and full revivals. Exciting the EMP WP with a
single voltage pulse results in a rather broad distribution of
wave numbers in the EMP WP, so that the analysis of the revival
structure, which is based on a sharply-peaked distribution of
wave numbers around $\tilde M$, is certainly invalid.

With the drift velocity deduced from $R$ and $T_{\text{cl}}$, we
estimate the thermal length to be $\zeta_{\text{T}} \approx 3\rm \mu
m$ which is smaller than $\zeta_{\text{G}}$. The experiment therefore
was in a regime where thermal noise dominates. For a good
signal-to-noise ratio, the requirement $u_0 \gg 2$~mV had to be
satisfied; this criterion was met in the experiment under
consideration. The life time of the wave packet as deduced from a
calculation outlined in this section is $\sim 1000 \, T_{\text{cl}}$.
Therefore, the rapid decay of the signal cannot be attributed to
dissipation into the phonon system.

\subsection{Non-polar Semiconductors: Si and Ge}

In non-polar semiconductors, piezoelectric coupling is absent, and the
rate of phonon emission by the plasmons is suppressed for the
long-wavelength plasmons typically involved. In this case we find that
the ratio of the life time and revival time,
\begin{equation}
\frac{\Gamma_{\tilde M}^{-1}}{t_{\text{rev}}} \sim \left[
\frac{R}{\ln R} \,\,\, \frac{1}{\tilde{M}^2} \right]^2 \quad ,
\end{equation}
can be made much larger than unity by adjusting the size of the QH
sample. For a millimeter-size sample at typical magnetic fields
($2\pi R \approx 1$~mm, $B=10$~T), the ratio
$\Gamma_{\tilde M}^{-1}/ t_{\text{rev}} \sim 1$~($25$) for $\tilde M
= 50$~($20$). The revival structure of EMP WP disussed in
Sec.~\ref{sec3} should therefore be observable for samples with  
non-polar semiconductor host materials, {\it e.g.}, in
Si/Ge-heterostructures.\cite{sigehetero}

\section{Summary and Conclusions}\label{sec5}

In this paper, we have examined the formation and evolution of
edge-magnetoplasmon wave packets in nanostructures. These
wave packets are formed as superpositions of edge magnetoplasmons
that are the only low-lying excitations in finite quantum-Hall
samples. By using a sequence of short pulses in the excitation
process, it is possible to produce a superposition with mode numbers
sharply peaked around a central value $\tilde M$. We have shown that
for such wave packets the initial motion is periodic with a period
$T_{\text{cl}}$. After several of these cycles, the wave packet
collapses and a sequence of fractional and full revivals commences.
This revival structure is analogous to that of Rydberg wave packets
in atomic systems; its relevant time scale is the revival time
$t_{\text{rev}}$.

We find that experiments that use capacitive coupling to the
charge-density fluctuation that is associated with the EMP WP both for
creation and detection of the EMP WP are analogous to Rydberg
wave-packet experiments that use the phase-sensitive Ramsey method of
detection. In both types of measurement, the semiclassical motion as
well as the revival structure is seen in the time variation of the
envelope function of a rapidly oscillating signal.

We have shown that thermal effects have no influence on the 
propagation of the wave packet. Examining possible scenarios for
energy loss of EMP, we found that plasmon-phonon coupling due
to piezoelectric effects (in polar semiconductors such as GaAs)
causes the wave packet to decay with a life time that is typically
less than the revival time. However, for 2D ES fabricated in
non-polar semiconductors such as Si or Ge, piezoelectric coupling
is absent and it is possible to produce wave packets with large
values of $\tilde M$ that will evolve for times of order
$t_{\text{rev}}$ without appreciable decay. In this way, it should be
feasible to detect fractional revivals in experiments.
The analysis given in this work of a previous
experiment\cite{ray:prb:92} that examined the classical motion of EMP
WP only can serve as a guideline for future experimental studies of
the EMP WP revival structure.

\acknowledgements{
We thank R.~C.\ Ashoori, D.\ DiVincenzo, S.~M.\ Girvin, S.~A.\
Mikhailov, J.~J.\ Palacios, D.\ Pfannkuche, and R.\ Haussmann for
stimulating discussions. This work was funded in part by NSF grants
No.~DMR-9416906 and PHY-9503756. U.Z.\ gratefully acknowledges
financial support from Studienstiftung des deutschen Volkes (Bonn,
Germany).
}

\widetext
\pagebreak
\begin{appendix}
\narrowtext
\noindent

\section{Possible Excitation Scenarios for the Initial Wave Packet}
\label{excite}
\subsection{Single Short Pulse}
In the experiments on $\mu$m--size quantum dots\cite{ray:prb:92} by 
Ashoori {\it et al.} a single short pulse was applied to prepare the
initial wave packet. In our model, the corresponding pulse
characteristics is
\begin{equation}
u(t) = \left\{ 
\begin{tabular}{cc}
$u_0$ & \,\, \mbox{ for $0\le t \le T_{\text{exc}}$} \quad , \\
0 & \,\, \mbox{ otherwise} \quad .
\end{tabular}
\right.
\end{equation}
It is straightforward to calculate the field $B_{\pm M}(t)$ for this
case. We find
\widetext
\top{-2.8cm}
\begin{equation}
B_{\pm M}(t) = \left\{
\begin{tabular}{cc}
0 & \, \, $t\le 0$ \quad , \\
$(\nu M)^{1/2} \, V^{\text{ext}}_{\pm M} \, \frac{\mp 2 i 
u_0}{\varepsilon^{\text{C}}_{M}} \, \exp{\left[ \pm i 
\frac{\varepsilon^{\text{C}}_{M} t}{2} \right] } \,\, 
\sin{\left[ \frac{\varepsilon^{\text{C}}_{M} t}{2} \right] }$ & \, \,
$0 < t < T_{\text{exc}}$ \quad , \\
$(\nu M)^{1/2} \, V^{\text{ext}}_{\pm M} \, \frac{\mp 2 i 
u_0}{\varepsilon^{\text{C}}_{M}} \, \exp{\left[ \pm i 
\frac{\varepsilon^{\text{C}}_{M} T_{\text{exc}}}{2} \right] } \,\,
\sin{\left[ \frac{\varepsilon^{\text{C}}_{M} T_{\text{exc}}}{2}
\right] }$ & \, \, $T_{\text{exc}} \le t$
\quad .
\end{tabular} \right.
\end{equation}
\bottom{-2.7cm}
\narrowtext
\noindent
This type of excitation cannot lead to an EMP WP state with a sharply
peaked mode distribution and is unlikely to produce well-resolved
revivals.

\subsection{Multiple Short Pulses}\label{laser}
In analogy with the excitation of Rydberg wave-packet states in atoms
by laser pulses, we propose exciting the edge of a QH system using a
series of $N$ short pulses each of duration $T_{\text{exc}}$. For the
specific case of sinusoidal individual pulses this would give 
\begin{equation}
u(t) = \left\{ 
\begin{tabular}{cc}
$u_0 \, \sin\left[ \frac{2\pi}{T_{\text{exc}}} t \right]$ & \,\, for
$0\le t \le N \, T_{\text{exc}}$ \quad , \\ 0 & \,\, otherwise \quad .
\end{tabular} \right.
\end{equation}
In this case we find that for $t \ge N T_{\text{exc}}$
\widetext
\top{-2.8cm}
\begin{equation}
B_{\pm M}(t) = (\nu M)^{1/2} \, V^{\text{ext}}_{\pm M} \, \frac{4\pi
u_0}{T_{\text{exc}}} \, \exp \left[ \pm i \frac{
\varepsilon^{\text{C}}_{M} N T_{\text{exc}}}{2} \right] \,\,
\frac{\sin \left[\varepsilon^{\text{C}}_{M} N T_{\text{exc}} /2
\right]}{\left( \varepsilon^{\text{C}}_{M}\right)^2 - \left(\frac{2
\pi}{T_{\text{exc}}} \right)^2 } \quad ,
\end{equation}
\bottom{-2.7cm}
\narrowtext
\noindent
which is sharply peaked around a value $\tilde M$ satisfying
$\varepsilon^{\text{C}}_{\tilde M} = 2 \pi / T_{\text{exc}}$.
Using our notation from Sec~\ref{revistruc}, the EMP WP created with
the multiple-short-pulse technique satisfies $T_{\text{ph}}\equiv
T_{\text{exc}}$. The duration of the short pulses determines the
value of $\tilde M$, whereas the number of pulses $N$ determines the
width of the peak in $B_{\pm M}(t)$ at $\tilde M$.
Figure~\ref{lasexc} illustrates some possible distributions in $M$
that could be produced using this method. 

\subsection{Adiabatic Limit}
An important limit of our general results is the case of an
adiabatically varying potential $V^{\text{ext}}(t)$. In our formalism,
this corresponds to a pulse characteristics $u(t)$ which varies on a
time scale longer than the time scale set by the lowest EMP energy.
The time integral in expression Eq.~(\ref{field}) is then dominated by
the exponential, and $u(\tau)$ can be treated as a constant within the
range of integration. We find
\begin{equation}
B_{\pm M}(t) = (\nu M)^{1/2} \, V^{\text{ext}}_{\pm M} \, \frac{-
u(t)}{ \varepsilon^{\text{C}}_{M}} \, \exp{\left[ \pm i 
\varepsilon^{\text{C}}_{M} t \right]} \quad ,
\end{equation}
and the total density response $\tilde\varrho_{M}(t)$, derived from
Eq.~(\ref{sigma}), is
\begin{equation}\label{adiabat}
\tilde\varrho_{M}(t) = - u(t) \, V^{\text{ext}}_{M} \,\, \frac{\nu
M}{\varepsilon^{\text{C}}_{M}} \quad .
\end{equation}

The induced density in Eq.~(\ref{adiabat}) is the instantaneous
ground-state density that minimizes the energy in the presence of the
slowly-varying external potential. The energy in the presence of the
external potential can be expressed in terms of the charge density as
follows: 
\widetext
\top{-2.8cm}
\begin{equation}
E[\varrho] = \int d\theta \,\, \int d\theta' \,\, V^{\text{int}}(
\theta, \theta' ) \tilde\varrho(\theta) \, \tilde\varrho(\theta') +
u(t) \int d\theta \,\, V^{\text{ext}}(\theta)\, \tilde\varrho(\theta)
\quad .
\end{equation}
\bottom{-2.7cm}
\narrowtext
\noindent
The configuration that minimizes the energy functional is
\begin{equation}\label{variate}
\tilde\varrho_{M}(t) = - u(t)
\frac{V^{\text{ext}}_{M}}{V^{\text{int}}_{M}} = - u(t) \, 
V^{\text{ext}}_{M} \,\, \frac{\nu M}{\varepsilon^{\text{C}}_{M}}
\quad ,
\end{equation}
consistent with Eq.~(\ref{adiabat}). 

\section{Derivation of the Effective EMP--Phonon Coupling}
\label{plasphon}
In this section, the derivation of the effective coupling between the
EMP and 3D phonons is given. We start with the full Hamiltonian
[Eq.~(\ref{elecphon})] describing the interaction between 3D electrons
with the 3D lattice in the sample. It is convenient to study the
Fourier components of the 3D electron density
\begin{equation}\label{fourier}
\varrho^{\text{3D}}_{\vec q} = \int d^3 r \,\, e^{i \vec q \cdot
\vec r} \, \varrho^{\text{3D}}(\vec r) \quad .
\end{equation}
As we are dealing with a 2D ES which is confined, say, to the
$xy$-plane, we introduce the notation $\vec r = z \hat z +
\underbar r $ where $\hat z \perp \,\, \underbar r $ (in reciprocal
space: $\vec q = q_z \hat z + \underbar Q $, $\hat z \perp \,\,
\underbar Q $) and assume the electron density to be peaked strongly
at $z=z_0$: $\varrho^{\text{3D}}(\vec r) = \chi(z) \varrho^{\text{2D}}
(\underbar r )$. Then, the $z$-integration in Eq.~(\ref{fourier})
decouples from the rest of the 3D integral, and merely leads to a form
factor $F^{\perp}(q_z) = \int dz \, e^{i z \cdot q_z} \chi(z)$. We are
left with the 2D Fourier transform of the 2D electron density
$\varrho^{\text{2D}}(\underbar r )$.

\subsection{Disk Geometry}
Specializing to the case of a QH sample in the disk geometry (see
Fig.~\ref{geometry}), we can write approximately
\begin{equation}\label{densapprox}
\varrho^{\text{2D}}(\underbar r ) = \varrho_0 \,\, \Theta(R -
|\underbar r |) + \varrho^{\text{1D}}(\theta) \,\, \delta(R -
|\underbar r | ) \quad ,
\end{equation}
where $\varrho_0 = \nu / 2\pi$, $\Theta(x)$ is the Heaviside
step function, and $\theta$ is the coordinate along the edge (see
Fig.~\ref{geometry}). The remaining integrals can be performed. The
result is
\widetext
\top{-2.8cm}
\begin{mathletters}
\begin{equation}\label{3Ddens}
\varrho^{\text{3D}}_{\vec q} = \varrho^{\text{bulk}}_{\vec q} +
\varrho^{\text{edge}}_{\vec q} \quad ,
\end{equation}
\begin{equation}
\varrho^{\text{bulk}}_{\vec q} = F^{\perp}(q_z) \, \nu R^2 \, J_1(Q R)
\quad ,
\end{equation}
\begin{equation}
\varrho^{\text{edge}}_{\vec q} = F^{\perp}(q_z) \, \sum_{M>0} i^M \,
J_{M}(Q R)\, [\nu M]^{1/2} \,\, \left( b_{M}^{\dagger} \, e^{- i M
\theta_{\vec q}} + b_{M} \, e^{i M \theta_{\vec q}} \right) \quad ,
\label{edgedens}
\end{equation}
\end{mathletters}
\bottom{-2.7cm}
\narrowtext
\noindent
where we write $\theta_{\vec q}$ for the polar angle of the vector
$\underbar Q $ in the $xy$-plane, {\it i.e.}, the azimuthal angle of
$\vec q$ in 3D. We remind the reader that $\underbar Q $ is the
projection of the wave vector $\vec q$ onto the plane where the 2D ES
is located. To get Eq.~(\ref{edgedens}), we inserted
Eq.~(\ref{density}) for $\varrho^{\text{1D}}(\theta)$. Using
expression Eq.~(\ref{3Ddens}) for the 3D electron density
$\varrho^{\text{3D}}_{\vec q}$ in Eq.~(\ref{elecphon}), we find the
Hamiltonian Eq.~(\ref{coupli}) for the coupling between the EMP and
phonon modes, with
\begin{equation}\label{diskres}
C^{\lambda}_{M}(\vec q) = [\nu M]^{1/2} \, M^{\lambda}(\vec q)
\,\, i^M \, J_{M}(Q R) \,\, F^{\perp}(q_z)
\end{equation}
as the coupling strength. Note that $J_{M}(x)$ denotes the
$M^{\text{th}}$-order Bessel function of the first kind.

\subsection{Strip Geometry}
For the sake of completeness, we give the corresponding results for
the case of a QH bar (strip geometry). By QH bar, we mean a sample
with periodic boundary conditions applied in the $\hat x$-direction,
and open boundary conditions in the $\hat y$-direction. This
configuration space corresponds to the surface of a cylinder with axis
in the $\hat y$-direction. Although this geometry is not appropriate
for the observation of EMP WP revivals, it can be useful in
analyzing experiments in which edge disturbances travel along the
edge of a Hall bar.\cite{zhi:prl:93}

In analogy with Eq.~(\ref{densapprox}), we write for the case of the
strip geometry
\begin{equation}
\varrho^{\text{2D}}(\underbar r ) = \varrho_0 \,\, \Theta(W - y) +
\varrho^{\text{1D}}(\theta) \,\, \delta(y - W ) \quad ,
\end{equation}
with $W$ denoting the width of the strip. We again end up with an
expression like Eq.~(\ref{3Ddens}), and find for the Hamiltonian
describing the plasmon-phonon coupling:
\widetext
\top{-2.8cm}
\begin{equation}\label{stripcoup}
H^{\text{pl-ph}} = \sum_{M>0 \atop \vec q, \lambda} C(\vec q, \lambda)
\left( a_{\vec q, \lambda} + a^{\dagger}_{- \vec q, \lambda} \right)
\left( b_{M} \, \delta_{M, -R \cdot q_x} +  b^{\dagger}_{M} \,
\delta_{M, R \cdot q_x} \right)
\end{equation}
\bottom{-2.7cm}
\narrowtext
\noindent
The coupling strength is
\begin{equation}
C(\vec q, \lambda) = [\nu M]^{1/2} \, M^{\lambda}(\vec q) \,\,
F^{\perp}(q_z) \,\, e^{i W \cdot q_y} \quad ,
\end{equation}
which is different from the result Eq.~(\ref{diskres}) we found for
the disk case. Note that the factor $e^{i W \cdot q_y}$ is simply a
form factor describing the profile of the charge density in
$\hat y$-direction. Here, we have assumed a sharp confining potential
and therefore used a delta-function. In general, this is not the
experimentally realistic situation, and we will have to replace
$e^{i W \cdot q_y}$ by a form factor $F^{\parallel}(q_y)$. A similar 
form factor should in principle be included in the analysis for the 
disk geometry as well, but would not be important for small $M$. Note
the differences between the final expression for the coupling of EMP
to phonons, Eq.~(\ref{coupli}) for the disk geometry, and
Eq.~(\ref{stripcoup}) the for strip geometry.

\subsection{Specialization: Acoustic Phonons in Semiconductors}
For the physical situation we are concerned with in this paper,
acoustic phonons play the dominant r\^ole. In polar semiconductors,
as for instance in AlGaAs/GaAs heterostructures, phonon coupling
occurs due to both deformation-potential and piezoelectric scattering.
The bare 3D electron-phonon coupling reads
\begin{equation}
M^{\lambda}(\vec q) 
= \left( \frac{\hbar}{2\rho \Omega [\epsilon(Q)]^2 
\omega^{\lambda}_{\vec q} }\right)^{\frac{1}{2}} \left[ D \, q \, \,
\delta_{\lambda, l} + i \,\, e h_{\text{14}} \,\,
{\cal M}_{\lambda}(\hat q) \right] \, ,
\end{equation}
where $\rho$ and $\Omega$ denote the 3D bulk density and the sample
volume, respectively. We have also introduced the strengths of the
deformation potential ($D$) and the piezoelectric coupling
($h_{\text{14}}$). For more details on phonons in AlGaAs/GaAs
heterostructures as well as numerical values of the parameters, see
Ref.~\onlinecite{lyo:prb:88}. A general reference on scattering
mechanisms in metals and semiconductors is
Ref.~\onlinecite{scattering}. The dielectric function $\epsilon(Q)$
incorporates screening of the original electron-phonon interaction
due to many-electron effects. Here, the 2D ES is in the QH regime,
and there is no screening, therefore we set $\epsilon(Q) \rightarrow
1$. Finally, the functions ${\cal M}_{\lambda}(\hat q)$ model the
directional dependence of the piezoelectric phonon coupling. For a
2D ES which lies in the (100) plane of GaAs, we have\cite{lyo:prb:88}
\begin{mathletters}
\begin{eqnarray}
[{\cal M}_l(\hat q)]^2 &=& \frac{9}{2} \frac{Q^4 q_{z}^{2}}{q^6}
\quad , \\
\protect[{\cal M}_t(\hat q)]^2 &=& 2 \frac{Q^2 q_{z}^{4}}{q^6} +
\frac{1}{4} \frac{Q^6}{q^6} \quad ,
\end{eqnarray}
\end{mathletters}
for longitudinal and transverse modes respectively.  
\begin{figure}[hbt]
\epsfxsize=8.5cm
\centerline{\epsffile{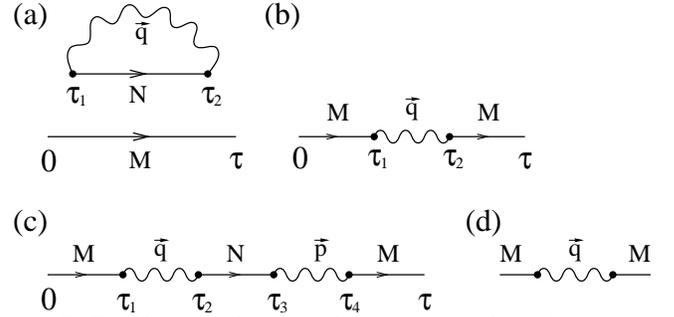}}
\caption{Diagrams involved in the calculation of the plasmon
propagator (a, b, c) and the plasmon self-energy (d), arising from
plasmon-phonon coupling. Straight lines denote the free plasmon
propagator ${\cal G}_{M}^{0}(\tau)$. The wavy line is the usual
phonon propagator ${\cal D}_{\lambda}^{0}(\vec q, \tau_1 - \tau_2)$.
(a)~Disconnected diagram; does not contribute to ${\cal G}_{M}(\tau)
$. (b)~Connected diagram; leading contribution to ${\cal G}_{M}(\tau)
$. (c)~Higher-order contribution to the plasmon propagator. Due to
azimuthal symmetry such a diagram vanishes unless $M=N$. The
self-energy insertion in this diagram is improper. The proper
self-energy is diagonal in angular-momentum indices and consists of
the single diagram shown in (d).}
\label{diagrams}
\end{figure}

\section{Plasmon Self-Energy due to Phonon Coupling}\label{selfen}
Due to the plasmon-phonon coupling, the plasmons acquire a
non-vanishing imaginary part of the self-energy that is related
to the rate of phonon emission/absorption by the plasmons. Here, we
calculate the plasmon self-energy using diagrammatic perturbation
theory. Due to azimuthal symmetry, the self-energy is diagonal in
angular-momentum indices. As the Hamiltonian of the coupled
EMP-phonon system is quadratic, the leading-order diagram gives
the exact result for the self-energy. Note that we consider here the
coupling of chiral 1D plasmons to 3D phonons; the problem of chiral
1D plasmons coupled to 1D phonons with implications for quantum-Hall
edges has been discussed previously.\cite{hei:prl:96}

The full Hamiltonian of the coupled EMP-phonon system (without the
external potential forming the initial wave packet) is
\begin{equation}
H' = H^{\text{pl}}_{0} + H^{\text{ph}}_{0} + H^{\text{pl-ph}} \quad ,
\end{equation}
where $H^{\text{pl}}_{0}$ is given by Eq.~(\ref{freeplasm}),
the expression Eq.~(\ref{coupli}) for $H^{\text{pl-ph}}$ in the disk
geometry has been derived in Appendix~\ref{plasphon}, and
$H^{\text{ph}}_{0}$ describes a system of free 3D phonons with
dispersion relation $\omega^{\lambda}_{\vec q}$:
\begin{equation}
H^{\text{ph}}_{0} = \sum_{\vec q, \lambda} \omega^{\lambda}_{\vec q}
\,\, a^{\dagger}_{\vec q, \lambda} \, a_{\vec q, \lambda} \quad .
\end{equation}
We want to calculate the single-plasmon Matsubara Green's function
defined in Eq.~(\ref{greensfunc}), which can be written as a sum over
all distinct connected diagrams. Three of the diagrams appearing in
${\cal G}_{M}(\tau)$ are shown in Fig.~\ref{diagrams}. All diagrams of
higher than second order are reducible. The sums over phonon wave
vectors that are implicit in the diagrams enforce the conservation of
the plasmon wave number in each higher-order diagram. Using the
standard definition\cite{mahan} for the phonon propagator
\widetext
\top{-2.8cm}
\begin{equation}
{\cal D}_\lambda^0 (\vec q, \tau_1 - \tau_2) = - \left< \mbox{T}_\tau 
\left[ a^{\dagger}_{- \vec q, \lambda}(\tau_1) + a_{\vec q, \lambda}
(\tau_1) \right] \left[ a^{\dagger}_{\vec q, \lambda}(\tau_2) +
a_{-\vec q, \lambda} (\tau_2) \right] \right>_0
\end{equation}
\bottom{-2.7cm}
\narrowtext
\noindent
we find that the full plasmon propagator is given exactly by
\begin{equation}
{\cal G}_{M}(i \omega_n) = \big[ i \omega_n - \varepsilon^{\text{C}}_M
- \Sigma_{M}(i \omega_n) \big]^{-1}
\end{equation}
with the plasmon self-energy 
\begin{equation}\label{selfres}
\Sigma_{M}(i \omega_n) = \sum_{\vec q, \lambda} \left|
C^{\lambda}_{M}(\vec q)\right|^2 \, {\cal D}_\lambda^0 (\vec q, i
\omega_n) \quad ,
\end{equation}
which is the contribution from second-order perturbation theory,
expressed diagrammatically in Fig.~\ref{diagrams}(d). The same result
for the plasmon propagator is obtained when integrating out the phonon
degrees of freedom in a path-integral expression for the partition
function of the coupled EMP-phonon system.

After continuation to real frequencies, it is possible to find the
imaginary part of the self-energy for the retarded Green's function
\begin{equation}
\Im m \, \Sigma_{M}(\omega) = \pi \sum_{\vec q, \lambda} \left|
C^{\lambda}_{M}(\vec q)\right|^2 \, \left[ \delta \big(\omega + 
\omega_{\vec q} \big) - \delta \big( \omega - \omega_{\vec q} \big)
\right] \, . 
\end{equation}
From this expression, we read off the damping rate for the mode with
wave number $M$ as expressed in Eq.~(\ref{damprate}).

\widetext
\end{appendix}


\begin{thebibliography}{10}
\bibitem{wiggles}
Note that the existence of the rapid oscillations necessarily follows
from the observability criterion for revivals: $t_{\text{rev}}\gg
T_{\text{cl}}$ implies $T_{\text{cl}}\gg T_{\text{ph}}$ (which is
equivalent to $\tilde M\gg 1$). The charge-density configuration of
the EMP WP consists of a finite classical wave train which has a large
dominant wave number $\tilde M$. In this it differs from a Rydberg
wave packet which is a complex wave.
\end{thebibliography}
\end{document}